\documentclass[aps,prb,superscriptaddress,a4paper, amsfonts, amssymb, amsmath, reprint, showkeys, nofootinbib, twoside]{revtex4-1}
\usepackage[english]{babel}
\usepackage{comment}
\usepackage[utf8]{inputenc}
\usepackage[colorinlistoftodos, color=green!40, prependcaption]{todonotes}
\usepackage[left=0.4in, right =0.4in]{geometry}
\usepackage{float}
\usepackage{graphicx,amssymb,amsmath,amsfonts,empheq}
\usepackage[pdftitle={TrappedExcitons}, pdfauthor={TorresLopez}]{hyperref} 
\usepackage{filecontents}
\usepackage{braket}

\makeatletter
\renewcommand*\env@matrix[1][\arraystretch]{%
  \edef\arraystretch{#1}%
  \hskip -\arraycolsep
  \let\@ifnextchar\new@ifnextchar
  \array{*\c@MaxMatrixCols c}}
\makeatother
\begin{document}
\title{Transition Metal Dichalcogenide Excitons in
Periodic Electrostatic Potentials:\\ Center-of-Mass Models}
\author{Jose M. Torres-L\'opez}
\affiliation{University of Texas at Austin, Austin TX 78712 USA}
\email[Correspondence email address: ]{josetorres@my.utexas.edu}%
\author{S.Kundu}
\author{Felipe H. da Jornada}
\author{Tony Heinz}
\affiliation{Stanford University}
\author{Allan H. MacDonald}
\affiliation{University of Texas at Austin, Austin TX 78712 USA}

\date{\today} 
\begin{abstract}
 Two-dimensional (2D) van-der-Waals materials are a promising platform for exciton state engineering. In this paper, we study the properties of
 excitons in 2D group VI transition-metal dichalcogenide (TMD) semiconductors that are modified by a periodic electrostatic potential through the quadratic Stark effect.  
 Using a model that retains only center-of-mass and valley degrees-of-freedom, we find that electrostatic potentials can drive optical valley splitting up to 
 $\sim 10 {\rm {meV}}$ and induce valley selective exciton dispersion.
 We explain why both properties are sensitive to the rotational symmetry of the electrostatic trapping potential using a combination of numerical results and analytical approximations. An important consequence of valley-splitting is that the lowest exciton band is non-degenerate and has a linear dispersion around $\boldsymbol{\gamma}$ that is expected to suppress thermal excitations, allowing true Bose condensation and superfluidity of excitons in two space dimensions.
 
\end{abstract}


\maketitle

\section{Introduction} 
\label{sec:one}
Two-dimensional electron systems realized by intentional stacking of van-der-Waals material layers have emerged as a flexible platform for engineered quantum correlated and topologically non-trivial phases of matter \cite{bistritzer2011moire,cao2018unconventional, cao2018correlated, wang2020correlated, behura2021moire, zhou2021half,lu2024fractional, zhai2025twistronics}.
The excitons of transition metal dichalcogenide (TMD) two-dimensional (2D) semiconductors \cite{mak2010atomically, qiu2013optical, brem2020tunable, wu2017topological, fogler2014high, wu2018theory, regan2022emerging, huang2022excitons, ugeda2014giant, du2023moire, shimazaki2021optical, smolenski2021signatures, salvador2022optical, upadhyay2026giant}, which are strongly bound, are particularly tunable. 
In this article we focus on how the exciton properties of 
2D TMD semiconductors can be altered by applying periodic electrostatic potentials \cite{shimazaki2021optical, smolenski2021signatures, thureja2022electrically, hu2024quantum, thureja2024electrically, upadhyay2026giant}.
Since electrostatic potentials shift conduction and valence bands in the same direction, they have no influence on band gaps or exciton binding energies in the smooth potential limit. However, gradients in the electrostatic potential polarize the relative motion of electrons and holes and yield local quadratic Stark shifts proportional the square of the potential gradient.  By this mechanism they impose a periodic effective potential on charge neutral excitons.  \\
2D TMD excitons have a pseudospin-1/2 valley ~\cite{cao2012valley, zeng2012valley, mak2012control, xiao2012coupled, yu2014valley, kim2014ultrafast, sie2017valley} degree-of-freedom.
The two valleys are coupled by e-h exchange interactions that can drive valley depolarization \cite{yu2014valley} and decoherence \cite{jones2013optical, hao2016direct}. 
In bulk 2D materials the valley states are degenerate at center-of-mass 
(CM) momentum $\textbf{Q}=0$, but split into linearly and quadratically dispersing branches at finite $\textbf{Q}$ \cite{andreani1990exchange, yu2014valley, wu2017topological, cudazzo2016exciton, ghazaryan2018anisotropic}.  Only $\textbf{Q}=0$ exciton states contribute to absorption, so the excitons states are valley-degenerate when probed optically, a property that we will refer to as optical degeneracy. Control of optical valley-splitting is 
desirable for versatile access to the exciton valley degree-of-freedom.
Previous work proposed the use of strain \cite{yu2014dirac} or external magnetic fields \cite{wu2017topological, salvador2022optical, glazov2025long-range} as mechanisms to induced an optical exciton gap. However, a complete understanding of the optical exciton spectra in the presence of general electrostatic potentials is lacking. We show that, similarly to the effect of strain, periodic electrostatic potentials can lift the optical degeneracy if they lack rotational symmetry $C_n$ ($n>2$), providing a mechanism for valley-selective exciton dispersion. Our numerical results are interpreted in terms of perturbation theory of the exchange coupling near $\mathbf{Q}=0$ as well as exact results for a toy model based on anisotropic harmonic potentials.\\
Our paper is organized as follows.  In Section~\ref{sec:two} we outline the general theory of excitonic states in a smooth periodic electrostatic potential. In Section~\ref{sec:four} we explain how optical valley degeneracy can be protected by in-plane rotational symmetry.  In Section~\ref{sec:three} we specialize to the case of electrostatic potentials produced by interdigitated gates that yield a set of parallel exciton confinement lines. This low symmetry case is relatively easy to realize experimentally \cite{thureja2022electrically, hu2024quantum} and always lifts optical exciton degeneracy; similar anisotropic effects have been studied in the context of strained monolayers \cite{yu2014dirac} and inorganic nanowires \cite{folie2020effect, weiss2021influence}. The dispersions of the exciton state energies are linear and quadratic parallel and perpendicular to the confinement lines, with opposite rapid dispersion directions for the low and high energy exciton states.  Finally, in Section~\ref{sec:five} we summarize our results and present our conclusions.


\section{Excitons in a Periodic Electrostatic Potential}
\label{sec:two}
\subsection{Exciton Potentials and Bands}
Excitons are collective excitations of semiconductors or insulators - bound states of conduction band electrons and valence band holes that appear at energies below the thermodynamic gap.  Because they are composite 
particles, excitonic states diagonalize two-particle 
Hamiltonians that have a discrete set of bound states
for each center-of-mass momentum $\mathbf{Q}$.  
In this paper we are interested in how the spectrum associated with the lowest energy excitonic bound state in TMD 2D semiconductors is altered by a periodic electrostatic potential, $V(\mathbf{r})$.

Provided that the potential gradients $e\mathbf{E}(\mathbf{r})$ generated by $V(\mathbf{r})$ are weak compared to the ratio of the exciton binding energy
to the exciton size ({\it i.e.} small compared to 
$\sim 0.3$ eV nm$^{-1}$ in 2D TMDs), the influence of an electric field is captured by adding a local quadratic stark shift to the exciton center-of-mass Hamiltonian.  The resulting potential-energy for excitons: $\Delta(\mathbf{r}) = -\alpha|\mathbf{E}
(\mathbf{r})|^2/2$ where $\alpha$ is the polarizability of the excitonic ground state. We leave the case of larger local electric fields, which requires a full fermionic electron-hole calculation; such a calculation was applied recently to the case of interlayer excitons by Ref. \cite{zhang2025engineering}, which are easily polarized beyond the perturbative regime due to their weaker binding.
Since $\alpha \approx 20 {\rm eV} {\rm nm}^2{\rm V}^{-2}$ for monolayer TMD semiconductor excitons \cite{pedersen2016exciton}, the maximum transverse electric fields for which the quadratic Stark shift model can be applied are $\sim 0.1 \, {\rm V}{\rm nm}^{-1}$ and the maximum possible local values for $\Delta(\mathbf{r})$ are $\sim 200$ meV. 


It is useful to start by neglecting the particle-hole exchange interactions, 
even though these have a large influence on exciton dispersion as we emphasize below.
The exciton Hamiltonian is then
\begin{align}
H_0 &= \left( -\frac{\hbar^2 \mathbf{\nabla}^2}{2M} + \Delta(\mathbf{r}) \right)\tau_0 
\end{align} 
where $M$ is the exciton mass and $\tau_0$ is the identity matrix in valley space.
When $\Delta(\mathbf{r})$ is periodic, the exciton states form Bloch bands, but remain valley-degenerate:  
\begin{equation}
\label{eq:pwexpansion}
|n,{\mathbf{Q}},v\rangle = \sum_{\mathbf{G}} c^n_{\mathbf{G}}(\mathbf{Q}) \, |\mathbf{Q}+\mathbf{G},v\rangle,
\end{equation}
where $n$ is a band index, $\mathbf{Q}$ is a wavevector in the Brillouin-zone
defined by the periodic potential, 
$\mathbf{G}$ is a reciprocal lattice vector, and 
$|\mathbf{Q},v\rangle$ is a valley $v$ free-exciton state with center-of-mass
momentum $\mathbf{Q}$.  We omit the valley index $v$ in $c^{n,v}_{\mathbf{G}}$ because the COM states are valley independent.  The expansion coefficients in Eq.~\ref{eq:pwexpansion} are normalized: $\sum_{\mathbf{G}} |c^n_{\mathbf{G}}(\mathbf{Q})|^2=1$. 
\begin{figure}
    \centering \includegraphics[width=45mm]{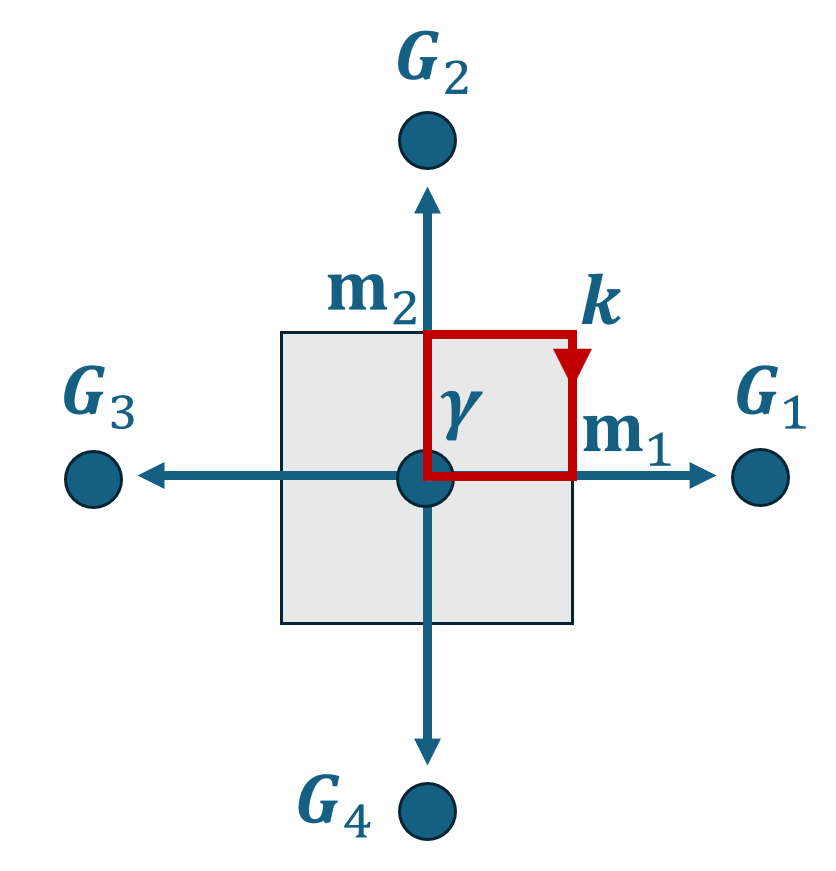}
    \hfill \includegraphics[width=45mm]{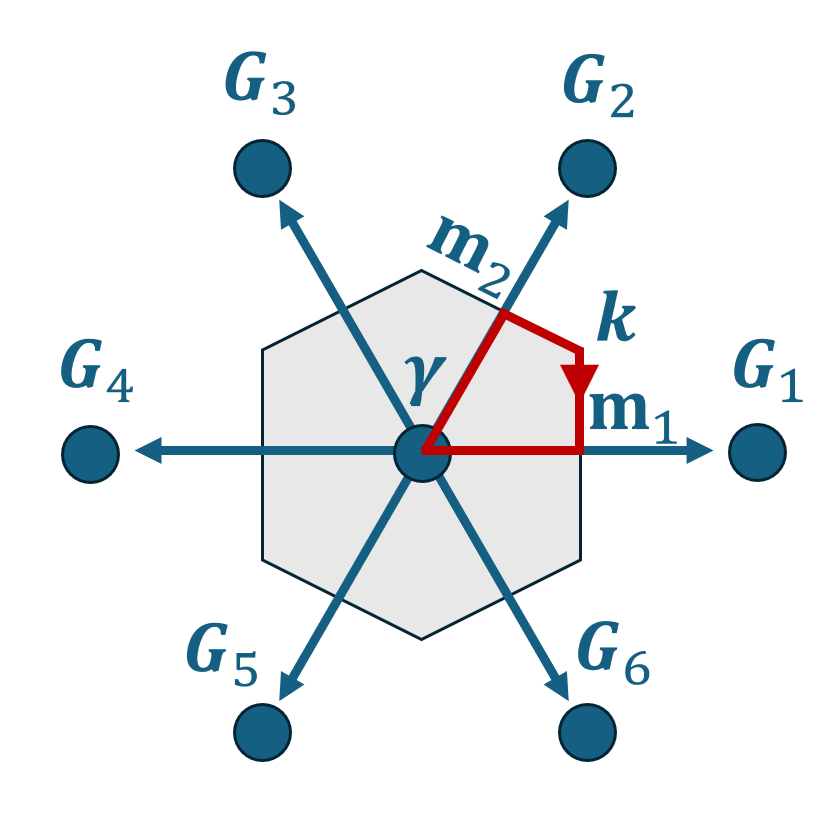}
    \includegraphics[width=44mm]{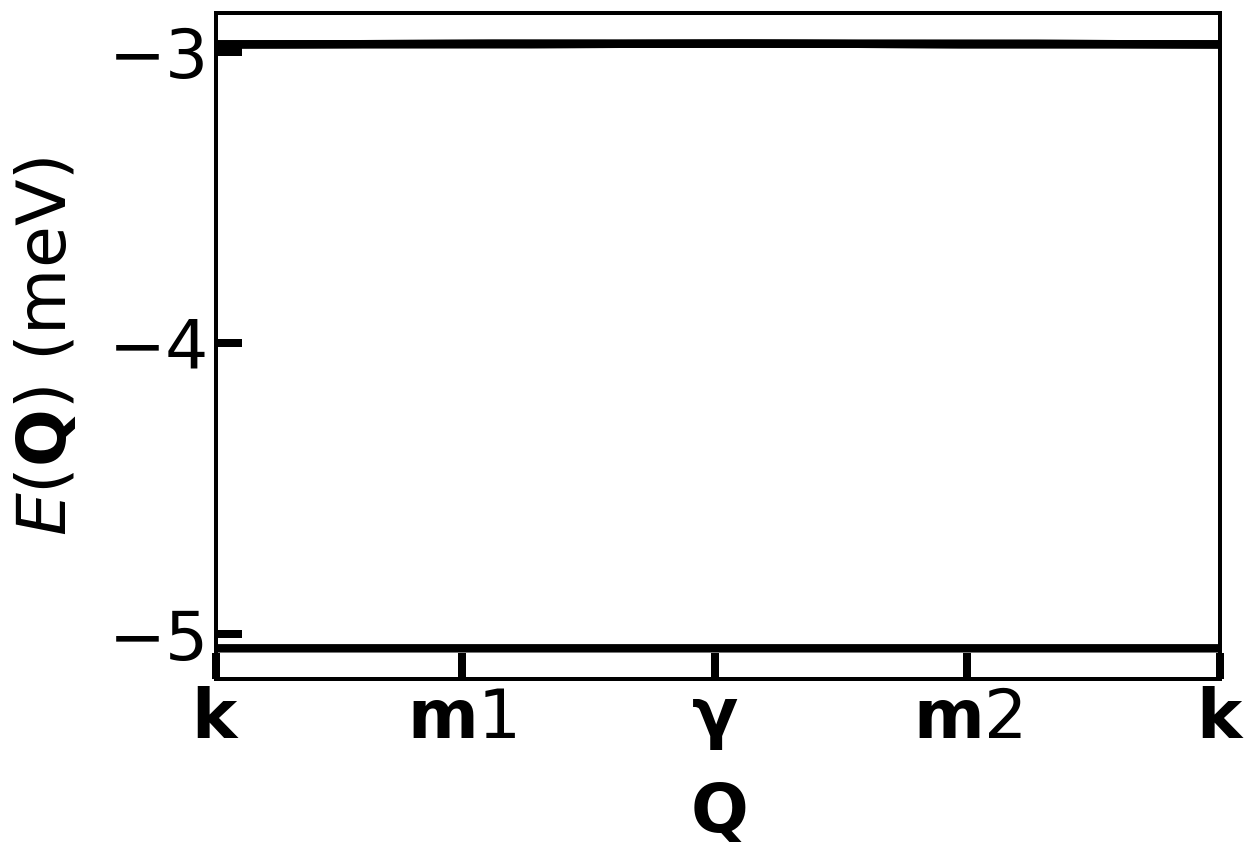}
    \hfill \includegraphics[width=46mm]{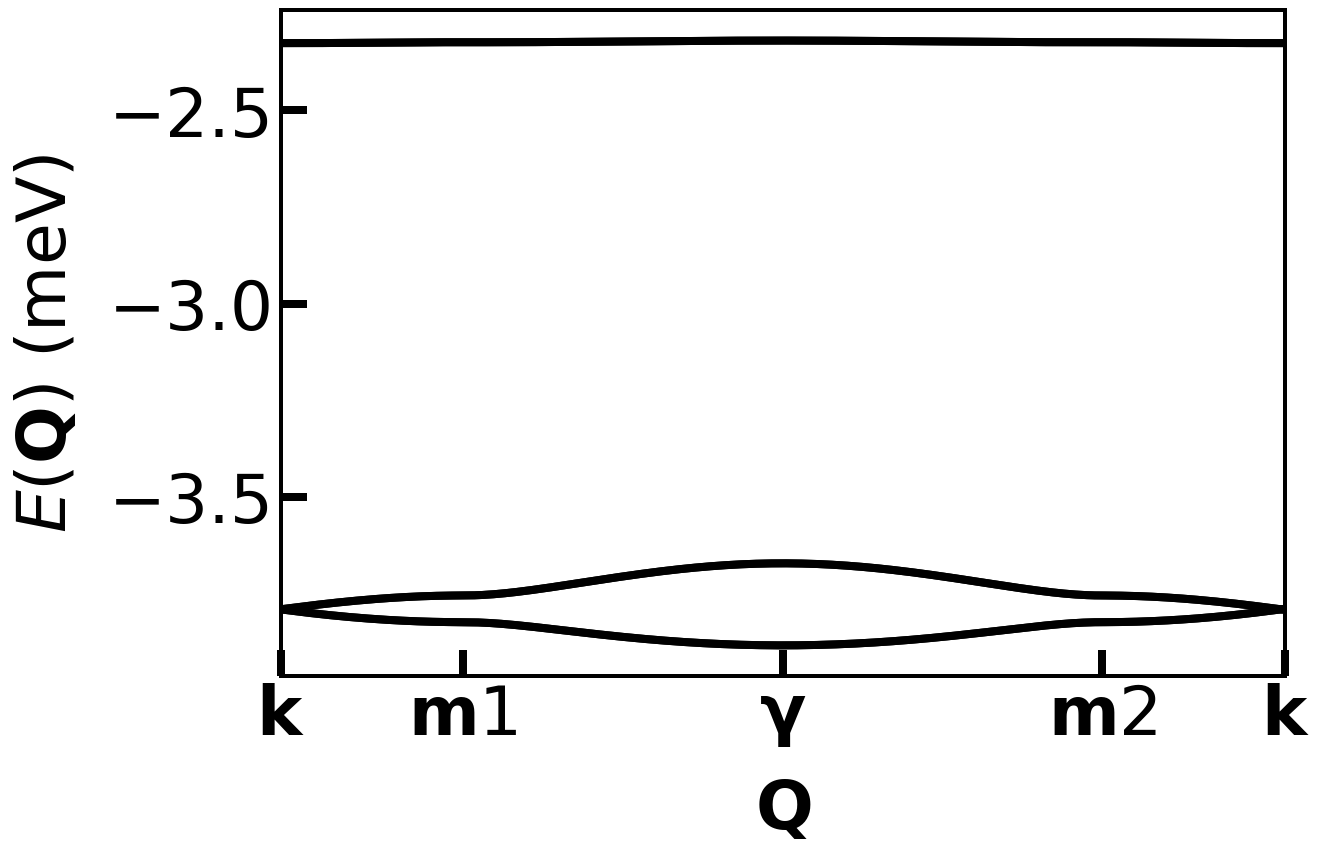}
    \caption{Top: Vectors in the first shell of the reciprocal lattice (blue arrows) and Brillouin zone (BZ) (grey area) for the triangular (left) and square lattices (right). Bottom: Six lowest exchange-free bands computed for $\Delta_j = 1$ meV on the first shell, exchange constant 
    $J = 0$ and lattice constant $a=50$ nm.  The states always occur in valley-degenerate pairs.
    Anticipating the anisotropy introduced later we have distinguished the BZ edge center ($m$) points as $\boldsymbol{m}_1, \boldsymbol{m}_2, ...$ following the notation for the first-shell $G_j$ vectors. 
    The bands are then plotted along the path $\boldsymbol{k}-\boldsymbol{m}_1-\boldsymbol{\gamma}-\boldsymbol{m}_2-\boldsymbol{k}$ illustrated in red. }
    \label{fig:G}
\end{figure}
A generic periodic exciton potential $\Delta(\mathbf{r})$, 
can be represented by its Fourier expansion with coefficients $\Delta_j$:
\begin{equation}
\Delta(\mathbf{r}) \approx \sum_{j=1}^N \Delta_j e^{i \mathbf{G}_j \cdot \mathbf{r}}.
\label{eq:delta}
\end{equation}
where $\mathbf{G}_j$ is a reciprocal lattice vector and $N$ is the number of reciprocal lattice vectors included in the expansion.  When they are present, point-group symmetries in the exciton potential $\Delta(\mathbf{r})$ impose constraints on the coefficients $\Delta_j$.
The first-shell of $\mathbf{G}_j$ vectors and the corresponding bands are 
illustrated in Fig.~\ref{fig:G} for triangular and square lattices, and the respective exciton potentials are shown on the left side of Figs. \ref{fig:C3Delta} and \ref{fig:C4Delta}. For Fig.~\ref{fig:G}, we have chosen a weak potential with $\Delta_j = 1$meV on the first shell of reciprocal lattice vectors 
and zero otherwise. In the remainder of the paper we will fix the value of this coefficient at $\Delta = 1$meV, corresponding to the strongest electrostatic potentials achieved experimentally. 
The electrostatic potentials we have in mind in this article have periodicities that can be achieved either lithographically \cite{forsythe2018band, thureja2022electrically, mrenca2023probing, yang2023gate, hu2024quantum, thureja2024electrically} or by forming domain patterns in adjacent 2D ferroelectrics \cite{rosenberger2020twist, weston2020atomic, andersen2021excitons}, and these cannot currently achieve lattice constants much below $\sim 50$ nm.  
As illustrated in Fig.~\ref{fig:G}, at this length scale 
even potentials with $\Delta_j = 1$meV in the first shell
produce flat exciton bands when the electron-hole exchange interaction is neglected.
The corresponding exciton wavefunctions are strongly localized near exciton potential 
maxima and the exciton states can be viewed as linear combinations of two-dimensional
harmonic oscillator states centered in different unit cells.

In the absence of exchange interactions, the exciton Hamiltonian in each valley is just that of a free particle of mass $M$ in a periodic potential, which has well-understood nearly free and strong potential limits.  In the nearly-free limit, the free particle dispersion is interrupted only near extended zone boundaries, whereas the strong potential limit produces 2D harmonic-oscillator-like states with exponentially small band widths.  The dividing line between these two limits occurs at potential strengths $\sim \hbar^2/Ma^2$ where $a$ is the periodic-potential lattice constant.  The bands plotted
in Fig.~\ref{fig:G} are in the strong potential limit in which the lowest energy exciton state in each 
valley is a linear combination of $1s$ harmonic oscillator state and the next two are linear combinations of $2p$ harmonic oscillator states.  In the square lattice case, the $p_x$ and $p_y$ exciton states are degenerate along the spaghetti plot lines.
\subsection{Exchange interactions}
The exciton Hamiltonian also includes an electron-hole exchange term due to the electric dipole moments created by coherence between electrons and holes.  
Since the exchange interaction does not break translational invariance, it produces a contribution to the exciton Hamiltonian that is diagonal in momentum $\mathbf{Q}$ and reciprocal lattice vector $\mathbf{G}$. (The exciton potential term on the other hand couples center-of-mass momenta that 
differ by reciprocal lattice vectors.) In 2D, the exchange Hamiltonian $H_X$ ~\cite{yu2014valley, glazov2014exciton}
is linear in center-of-mass wavevector,
\begin{widetext}
\begin{align}
\langle \mathbf{Q}+\mathbf{G'},v'|H_X|\mathbf{Q}+\mathbf{G},v\rangle &=  \delta_{\mathbf{G'},\mathbf{G}} \; J |\textbf{Q+G}| \; \Big[ \tau^0_{v',v} 
+  \cos(2\phi_\textbf{Q+G}) \tau^{x}_{v'v} + \sin(2\phi_\textbf{Q+G}) \tau^y_{v'v} 
\Big] \equiv \delta_{\mathbf{G'},\mathbf{G}} \, H^{J}_{\tau',\tau}(\mathbf{Q+G}) ,
\label{Eq:exchangeH}
\end{align}
\end{widetext}
where $\phi_\textbf{Q}$ is the orientation
angle of the 2D wavevector $\textbf{Q}$, and $\tau^{0,x,y}$ are Pauli matrices in valley space.  In the absence of an exciton 
potential, exchange leads to two valley-coupled modes with valley-states 
\begin{equation}
\ket{K_\pm} = \frac{1}{\sqrt{2}} [\ket{K} \pm e^{2i\phi_\textbf{Q}}\ket{K'} ] 
\end{equation}
and energies
\begin{equation}
E_\pm(Q) = \frac{\hbar^2 |\textbf{Q}|^2}{2M} + J|\textbf{Q}| \pm J|\textbf{Q}|.   
\end{equation}
These expressions imply that only one of the modes acquires an exchange-induced linear dispersion, while the other mode has an energy that is not changed by exchange and hence disperses quadratically. For this reason, the two valley modes are sometimes referred to as being exchange-like in the case of 
linear dispersion and Coulomb-like in the case of quadratic dispersion. 
At a 50 nm length scale for the periodic potential
$J|\textbf{G}| \sim 50$meV, larger than the kinetic-dispersion energy scale 
$\hbar^2 |\textbf{G}|^2/2M$, and the potential energy scale $\Delta_j$ by a factor 
of $\sim 10$. ($J \sim 0.4-0.6$ eV$\cdot$nm.)  The exciton band dispersion in the absence of exchange therefore plays a very small role in shaping the full exciton bands.  However, the wavefunctions of the exchange-free exciton bands do play a role as we now explain.\\

\subsection{Exciton Hamiltonian in the exchange-free Bloch band representation}
Insight into the role of exchange in the presence of a periodic 
electrostatic potential can be gained by calculating matrix elements of the exchange 
interaction in the representation of exchange-free exciton bands:
\begin{equation}
\label{eq:exchange}
    \langle m,\mathbf{Q}',\tau'|H^{J}|n,\mathbf{Q},\tau\rangle = \delta_{\mathbf{Q}',\mathbf{Q}} \;
    \sum_{\mathbf{G}} (c^m_{\mathbf{G}})^*(\mathbf{Q}) c^n_{\mathbf{G}}(\mathbf{Q}) \; H^{J}_{\tau',\tau}(\mathbf{Q}+\mathbf{G}). 
\end{equation}
In Eq.~\ref{eq:exchange} the labels $n,\mathbf{p},\tau$ refer to band, crystal momentum and valley. The sum on the right-hand-side of Eq.~\ref{eq:exchange} can be divided into two contributions, $\mathbf{G} = \mathbf{0}$ and $\mathbf{G} \neq \mathbf{0}$.  Note that only the intervalley matrix elements can give rise to valley selective properties, since the intravalley matrix elements are valley independent. To gain a qualitative understanding of valley-dependence in 
optical absorption and in the dispersion of optically excited excitons we focus on 
$\mathbf{Q}$ near zero where
\begin{widetext}
\begin{equation}
\label{eq:matrixelement}
    \langle m,\mathbf{Q},\tau'|H^{J}|n,\mathbf{Q},\tau\rangle \approx     (c^m_{\mathbf{0}})^*(\mathbf{0}) c^n_{\mathbf{0}}(\mathbf{0}) \;  \; H^{J}_{\tau',\tau}(\mathbf{Q})
    + \sum_{\mathbf{G}\ne 0} (c^m_{\mathbf{G}})^*(\mathbf{0}) c^n_{\mathbf{G}}(\mathbf{0}) \; H^{J}_{\tau',\tau}(\mathbf{G}).
\end{equation}
\end{widetext}
To understand optical absorption, we can set $\mathbf{Q} = \mathbf{0} \equiv 
\boldsymbol{\gamma}$ so that the first term on the right-hand side of 
Eq.~\ref{eq:matrixelement} vanishes.  
In the second term, the sum over $\mathbf{G}$ can be shown to vanish for
$\tau' \ne \tau$ for some symmetry-constrained cases. In particular, we consider the case of $C_n-$symmetric potentials ($n>2$) and assume that the exchange-free exciton states realize representations of $C_n$ with a definite angular momentum so that $c^m_{\mathbf{G}} = e^{il_m\theta_{\mathbf{G}}}c^m_{|G|}$ where $l_m \in \mathbb{Z}_n$ is defined only $\mod{n}$ because of the discrete rotational symmetry. 
Since the intervalley exchange matrix elements $H^{J}_{\bar{\tau},\tau}(\mathbf{G}) \propto e^{\pm 2i\theta_{\mathbf{G}}}$, inter-valley coupling occurs between bands $m'$ and $m$ only when $l_{m'}-l_{m} = \pm 2$ $\mod{n}$.
Intra-band inter-valley matrix elements always vanish for $n>2$.  
The exchange Hamiltonian therefore separates into decoupled blocks 
defined by subspaces of the form 
\begin{align}
    \begin{bmatrix}
        l+2, K' \\
        l, K
    \end{bmatrix} 
    &&  \begin{bmatrix}
        l-2, K \\
        l, K'
    \end{bmatrix}
\end{align}
where an entry $l, \tau$ represents the sector with angular momentum $l$ in valley $\tau$; the $l$ sector of the $K$ ($K'$) valley is coupled to $l+2$ ($l-2$) sector
in the opposite valley. 

These considerations suggest the following conclusion about the resulting exciton spectrum.  The $l$ sector in one valley is completely decoupled from the 
$l$ sector in the other valley.  As a result, 
valley degeneracy is preserved by the exchange interaction.
(It is always present in the exchange-free problem where it is 
guaranteed by time-reversal symmetry.) We emphasize that t
his conclusion relies on the fact 
that $l-l\pm 2 \neq 0 \mod n$, so that $C_2$ or lower symmetry
is a special case in which pronounced exchange-induced band gaps do occur.

Previous work pointed out that strain \cite{yu2014dirac} or magnetic fields \cite{wu2017topological, glazov2025long-range} can act respectively as in-plane or out-of-plane Zeeman fields, lifting the optical degeneracy.
Here we emphasize that low-symmetry periodic confinement can achieve the same result. In the following two sections, these principles are illustrated by sample computations and analytical approximations, starting with the comparison of periodic $C_n-$symmetric and weakly $C_n-$ breaking potentials for $n=3,4$ in the next section.
\section{Reduced Rotational Symmetry}
\label{sec:four}
\begin{figure}[H]
    \centering
    \includegraphics[width=45mm]{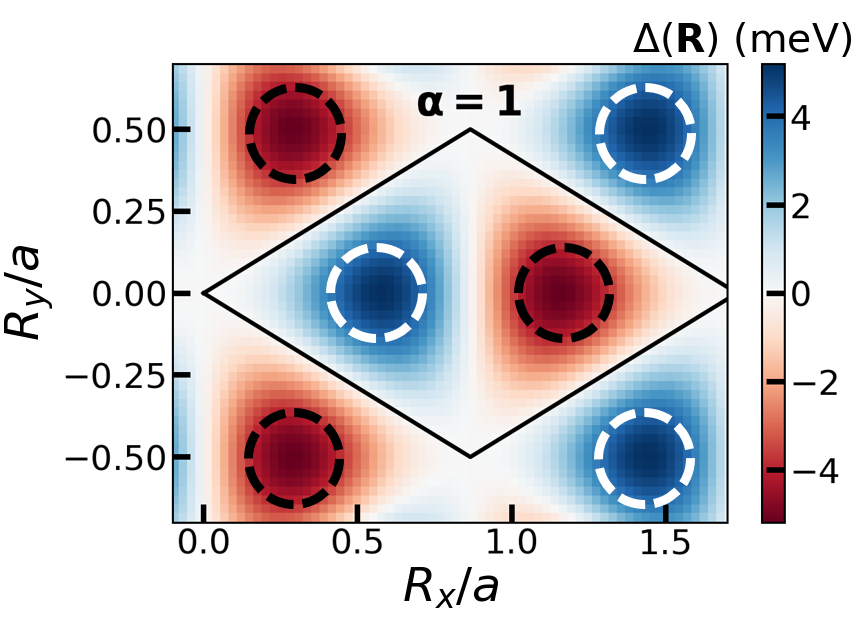}
    \hfill
    \includegraphics[width=45mm]{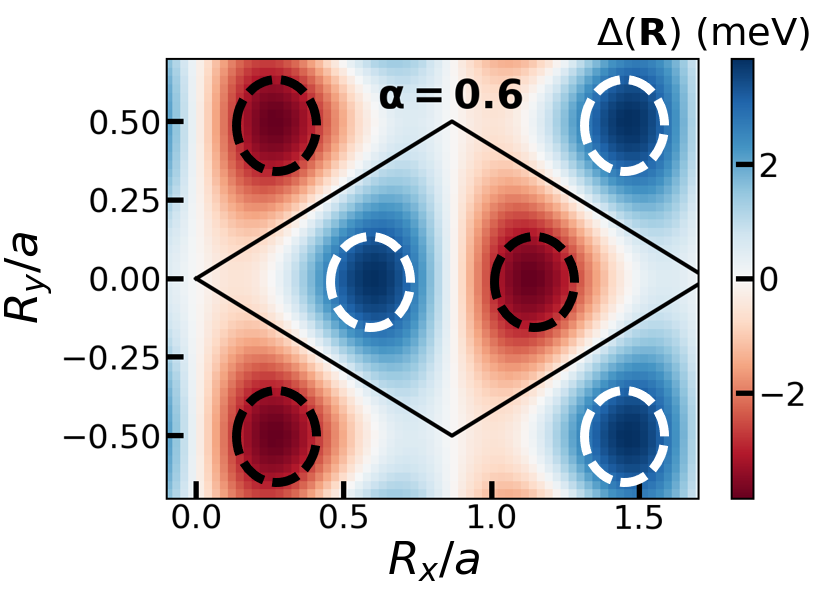}
    \caption{Triangular exciton potential: $C_3-$symmetric case with $\alpha = 1$ (left) and asymmetric case with parameter $\alpha = 0.6$ (right). For clarity, the unit cell is marked by the black parallelogram and the potential minima (maxima) are encircled by the black (white) dashed lines. The coordinate system has been shifted to place a 
    potential minimum at $a(2, 0)/\sqrt{3}$.}
    \label{fig:C3Delta}
\end{figure}

\begin{figure}
    \centering \includegraphics[width=45mm]{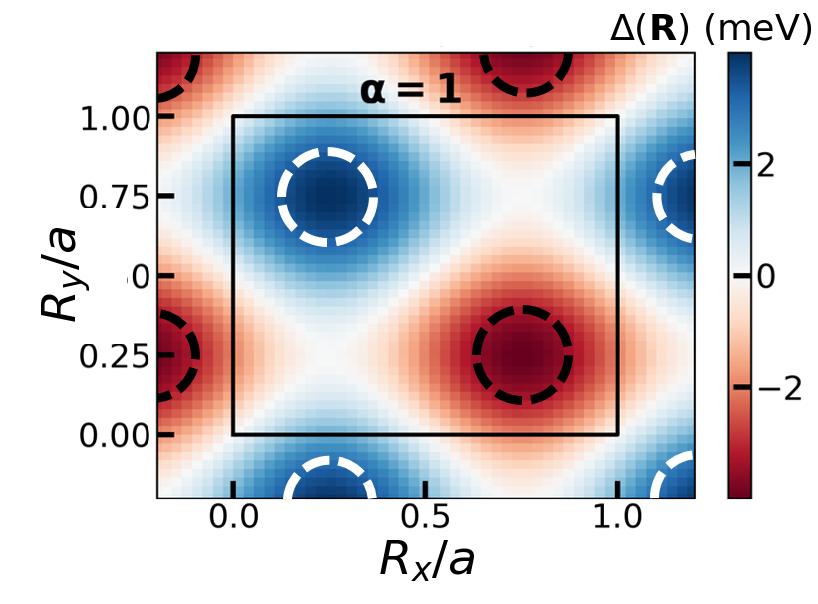}
    \hfill \includegraphics[width=45mm]{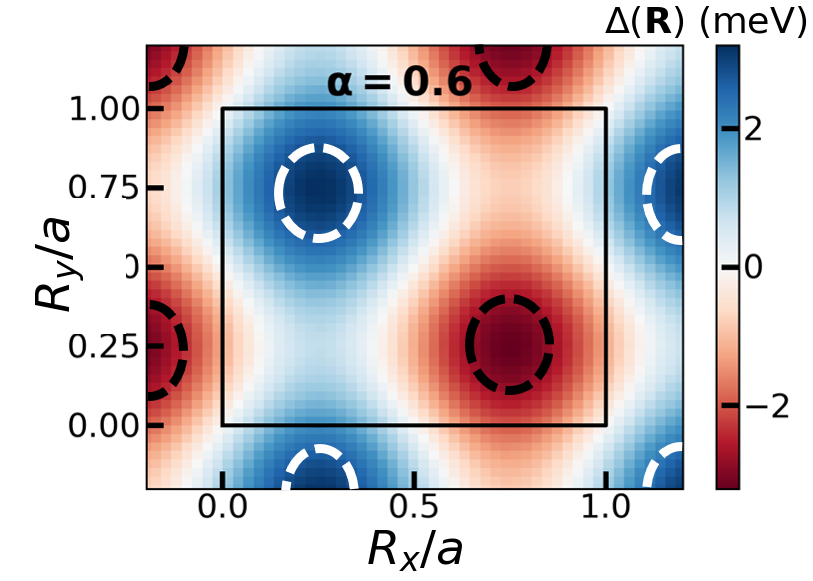}
    \caption{Square lattice potential for the $C_4-$symmetric case with $\alpha = 1$ (left) and for anisotropy parameter $\alpha = 0.6$ (right). The unit cell is marked by the black square, and the potential potential minima (maxima) are encircled by the black (white) dashed lines. The coordinate system has been shifted to place the potential minimum at $a(3/4,1/4)$.}
    \label{fig:C4Delta}
\end{figure}

\begin{figure}
    \centering \includegraphics[width=45mm]{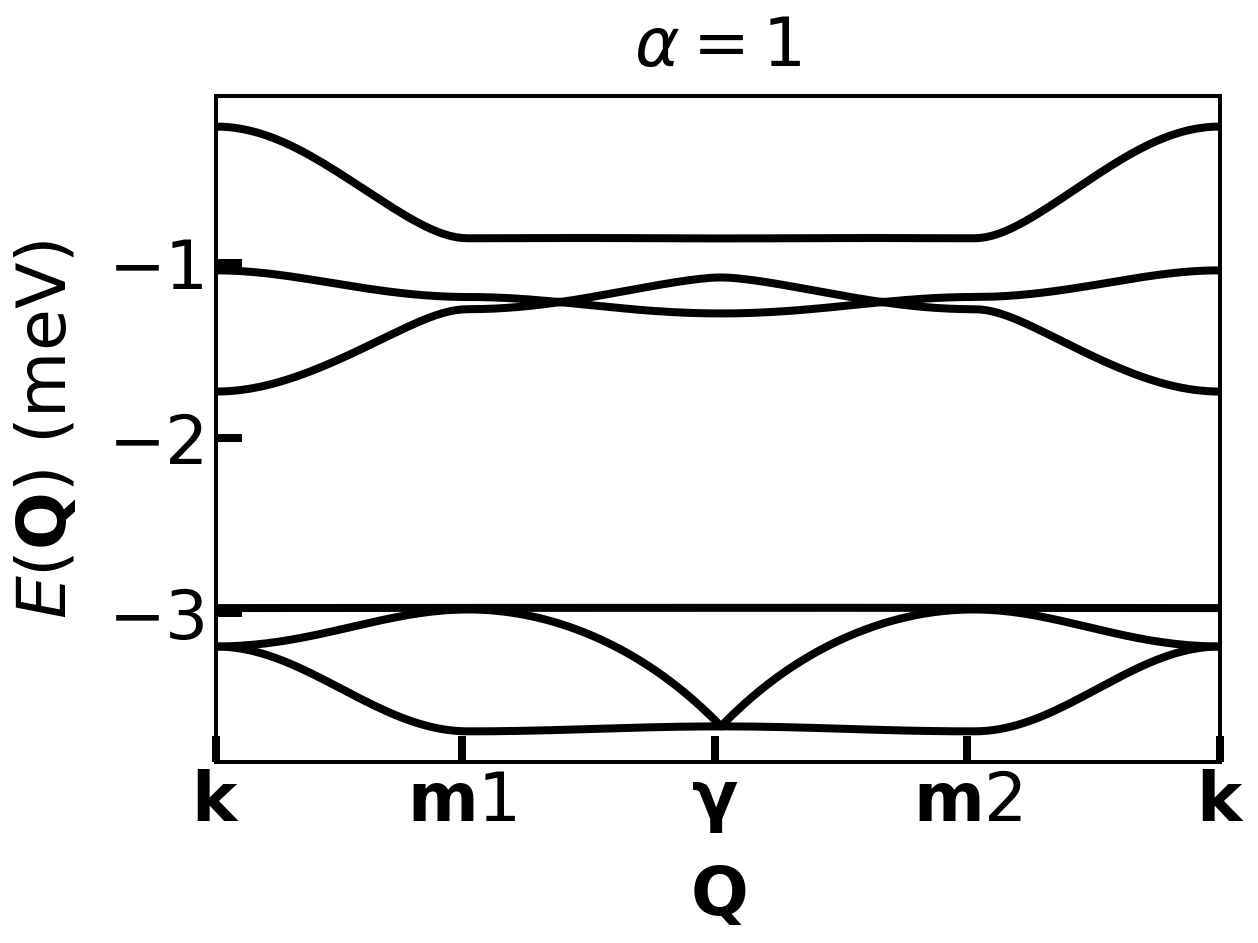} \includegraphics[width=45mm]{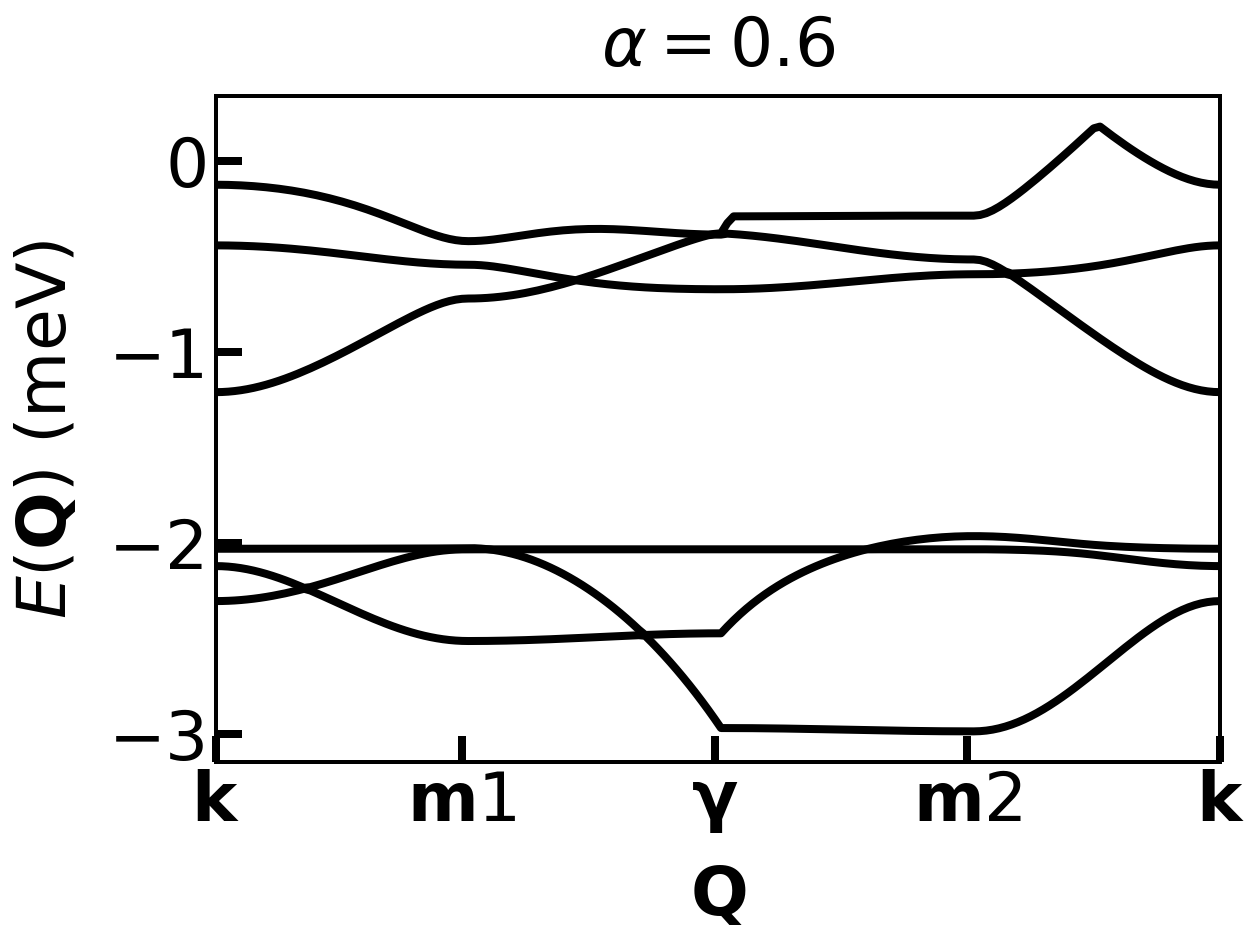} 
    \caption{Six lowest energy exciton 
    bands for the isotropic ($\alpha = 1$, left) and anisotropic ($\alpha = 0.6$, right) square lattices.}
    \label{fig:K4Delta}
\end{figure}

The numerical exciton bands are found by exact diagonalization of the full exciton Hamiltonian
\begin{equation}
    \begin{split}
    H = H_0 + H_X = \left( +\frac{\hbar^2 |\textbf{Q}|^2}{2M} + \Delta(\mathbf{r}) \right) \tau_0 \\
+ J |\textbf{Q}| [ \tau_0 +\cos(2\phi_\textbf{Q}) \tau_x + \sin(2\phi_\textbf{Q}) \tau_y ].
\end{split}
\end{equation}
where $\hbar \mathbf{Q}$ is the exciton momentum operator.
We study both triangular and square lattice models, and also 
distorted versions of these models in which the strength of $C_3$ and $C_4$ symmetry breaking is controlled by an anisotropy parameter $\alpha \in [0,1]$. 
For the triangular case, the potential harmonics are defined by $\Delta_1 = (\Delta_{4})^* =i$ and 
$\Delta_{2,6} = (\Delta_{3,5})^* = -\alpha i$ for $j=2,4,6$. We will compare the representative cases $\alpha = 0.6, 1$ with
the remaining model parameters fixed throughout this section at the values $a = 50$ nm and $J = 0.4 $ eV$\cdot$nm. 
For $\alpha = 0.6$, $\Delta_{C3}(\mathbf{r}) = - \sum_{j=1,2,3} 
2 |\Delta_j| \sin (\mathbf{G}_j\cdot \mathbf{r})$ has maxima and minima arranged on 
displaced triangular lattices as shown in Fig. \ref{fig:C3Delta}.
\begin{figure}[H]
    \centering \includegraphics[width=45mm]{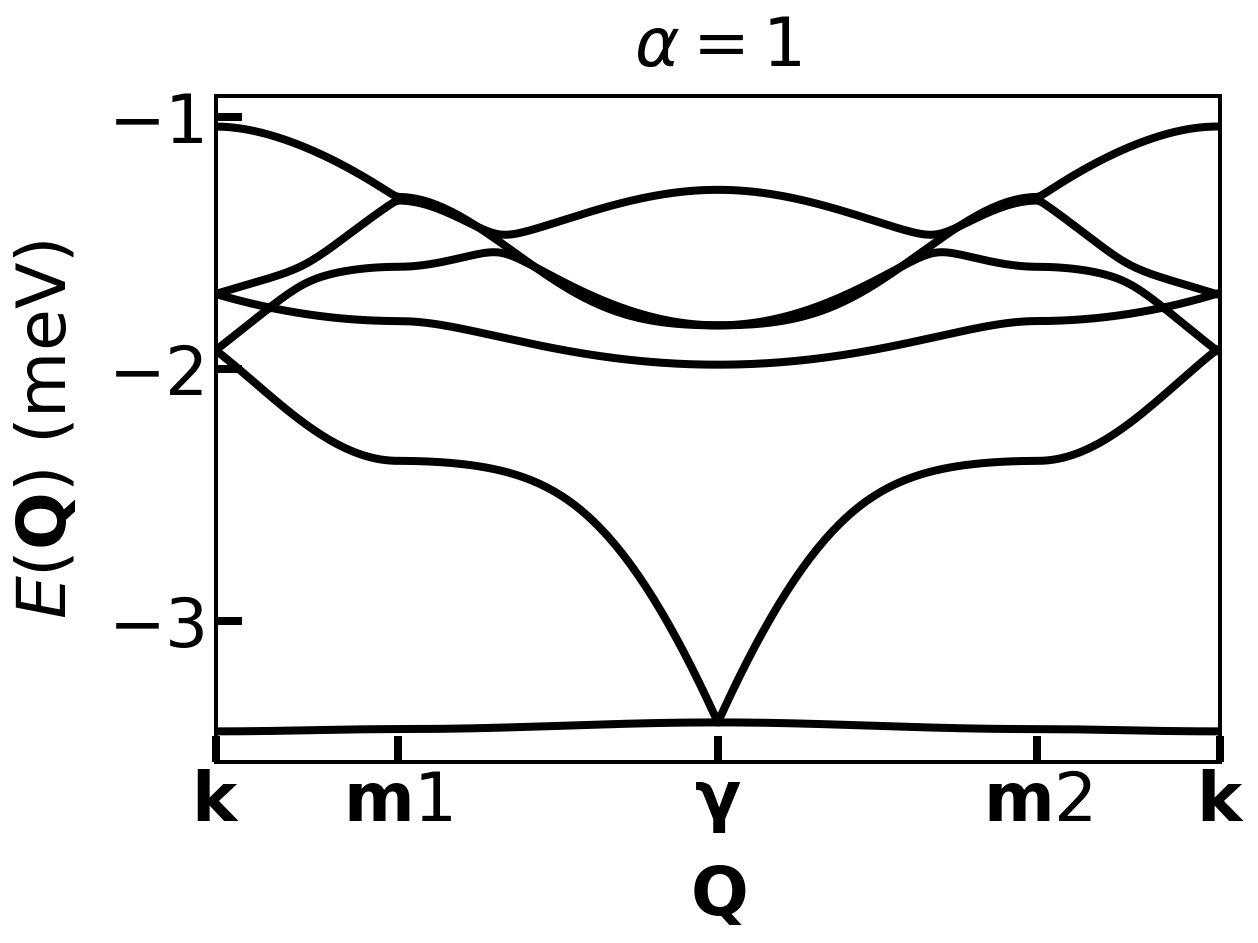} \includegraphics[width=45mm]{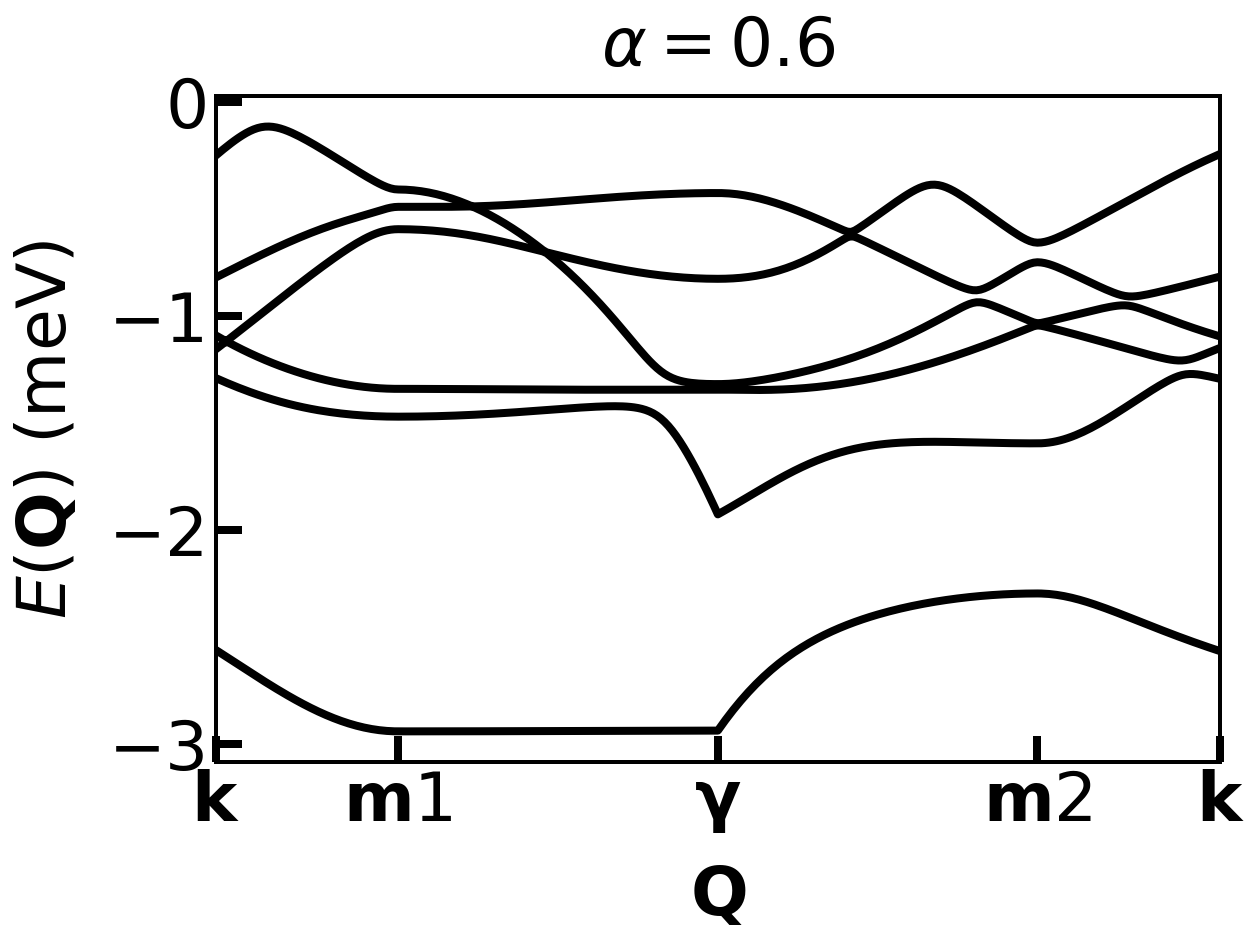} 
    \caption{
    The six lowest-energy bands for the isotropic ($\alpha = 1$, left) and anisotropic ($\alpha = 0.6$, right)  triangular lattice, plotted along the path defined in Fig. \ref{fig:G}.} 
    \label{fig:K3Delta}
\end{figure}
The results including exchange are plotted in Fig. \ref{fig:K3Delta} separately for $\alpha = 1$ (left) and $\alpha = 0.6$ (right). 
The $C_3$ bands on the left are approximately isotropic with 
linear and quadratically dispersing bands which are degenerate at $\boldsymbol{\gamma}$
as expected.  Over the plotted energy range, a single linearly dispersing band has avoided crossings with a series of flat bands whose energies are less influenced by exchange.  In contrast, the bands plotted on the right for the asymmetric $\alpha=0.6$ case are anisotropic, 
separated by a gap of about $0.7$ meV at $\mathbf{\Gamma}$, and shifted to higher energies relative to their counterparts on the left. We note that the two valley modes are now separated, and that each of the modes is linearly dispersive except in certain discrete mode-specific directions. The lowest energy band has linear exchange-induced dispersion along $Q_x$ but remains quadratic along $Q_y$, while the opposite is true for the second-lowest band. Therefore, the anisotropic bands cannot be categorized as having either ``exchange-like" or ``Coulomb-like" dispersion, in contrast to the isotropic case. We postpone a more detailed discussion of the anisotropic case to the following section, in which we consider a one-dimensional geometry that makes the distinction between exchange-like and Coulomb-like modes less nuanced.  The square lattice results in Fig.~\ref{fig:K4Delta} show the same qualitative features as the triangular case. The square potential is defined as $\Delta_{C4}(\mathbf{d}) = \sum_{j=1,2}2|\Delta_j|\cos (\mathbf{G}_j\cdot \mathbf{d})$ with $\Delta \equiv \Delta_{1} = (\Delta_{3})^* = 1i$ and $\Delta_{2} = (\Delta_{4})^* = \alpha\Delta_{1}$. The potentials corresponding to $\alpha = 1$ and $\alpha = 0.6$ are shown in Fig. \ref{fig:C4Delta}, and the corresponding band structures are illustrated in Fig. \ref{fig:K4Delta}.  

\begin{figure}
    \centering \includegraphics[width=60mm]{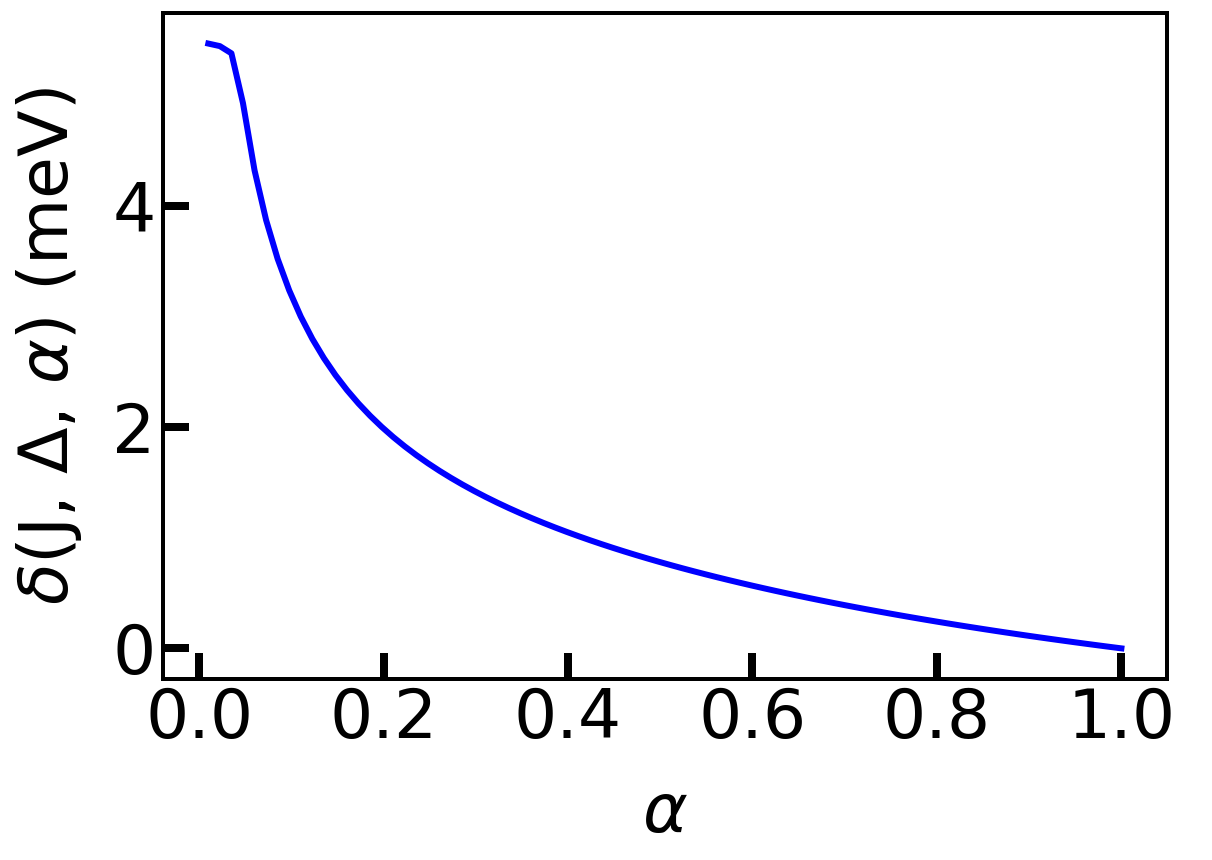}
    \caption{Magnitude of the exchange-induced valley gap $\delta(J,\alpha, \Delta) \equiv E^{n=1}_{\boldsymbol{\gamma}}(J,\alpha,\Delta)-E^{n=0}_{\boldsymbol{\gamma}}(J,\alpha,\Delta)$ for a square lattice as a function of anisotropy $\alpha$ for fixed $J= 0.4$ eV nm and $\Delta = 1$ meV.}
    \label{fig:GapA}
\end{figure}
\begin{figure}
    \centering \includegraphics[width=70mm]{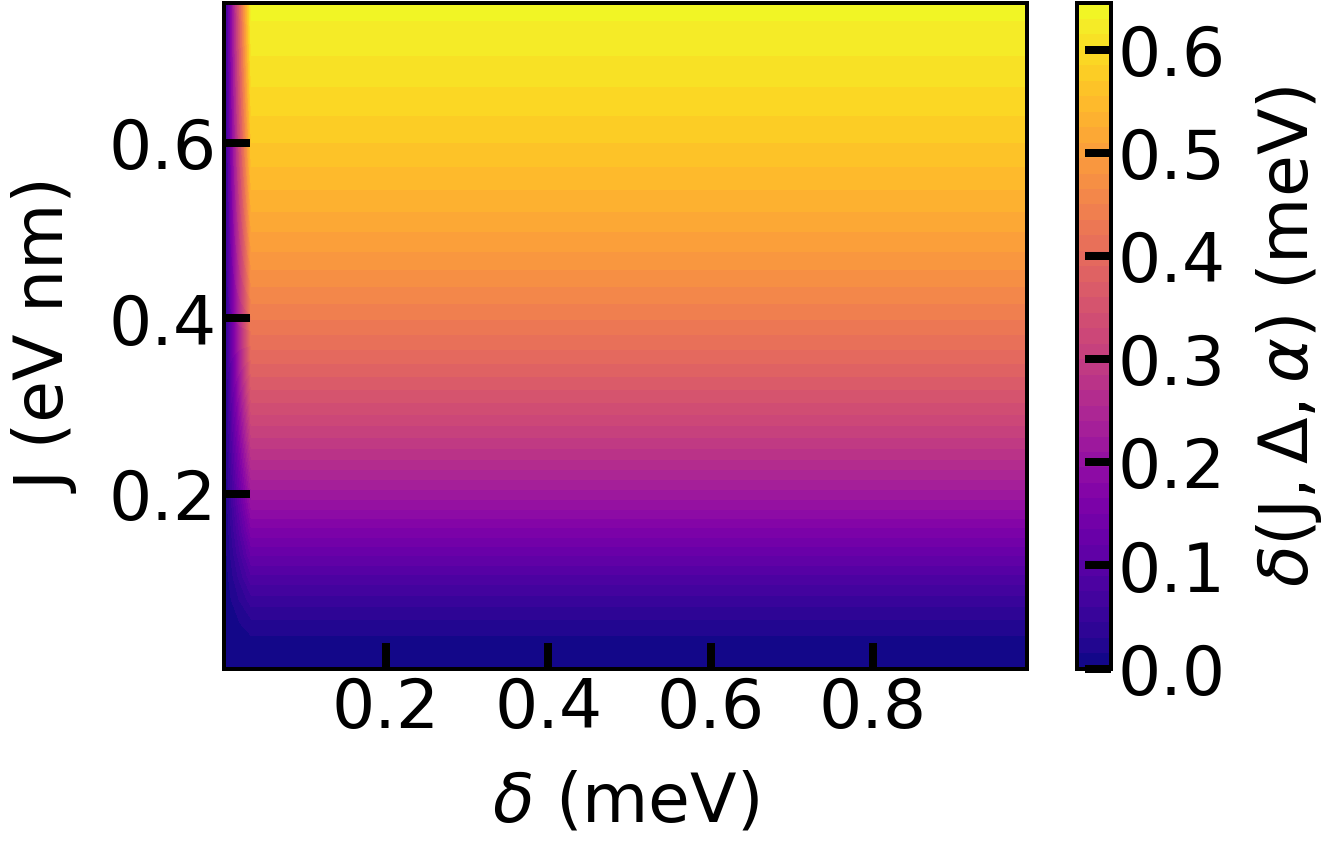}
    \caption{Magnitude of the valley gap $\delta(J,\alpha,\Delta)$ as a function of $J, \Delta$ for anisotropy parameter $\alpha=0.6$.}
    \label{fig:GapJD}
\end{figure}
The ground state optical gap at $\boldsymbol{\gamma}$ is a key parameter of exciton trap array systems. 
Its dependence on the values of the anisotropy parameter $\alpha$, the exchange parameter $J$, and the exciton potential parameter $\Delta_1$,
is illustrated for the square lattice case in Figs. \ref{fig:GapA} and \ref{fig:GapJD}.
We can understand these results qualitatively in terms of a tight-binding model for the exciton bands in the absence of exchange. Since the exciton trapping is parabolic close to potential extrema, the lowest energy band Wannier orbitals are 2D simple harmonic oscillator (SHO) ground states centered on the potential minima.
For the square lattice, $\Delta(r_x,r_y) = - 
V\left[ \cos(2\pi r_x/a)+ \alpha \cos(2\pi r_y/a)\right]$
has minima at $\mathbf{R}(n_x, n_y) = a[n_x, n_y]$, $n_{x,y} \in \mathbb{Z} $.  Expanding around the minima, the potential is that of an anisotropic oscillator
\begin{equation}
    \Delta_\mathbf{R}(r_x,r_y)  
    \approx  C 
     + 2V (\pi/a)^2 \left[ (r_x-R_x)^2 +  \alpha  (r_y-R_y)^2 \right]
\end{equation}
with constant shift $C \equiv -V (1+\alpha)$, spring constants $\kappa_x \equiv 4V(\pi/a)^2$, $\kappa_y= \alpha \kappa_x$ and frequencies $\omega_i \equiv (\kappa_i/M)^{1/2}$. Accordingly, the lowest energy exciton center-of-mass state near a minimum $\mathbf{R}$ has a 
Gaussian distribution in both real and momentum space.  It follows that their Wannier functions 
are 
\begin{equation}
    W_\mathbf{R}(\mathbf{r}) \approx \frac{1}{\sqrt{N_R}} e^{-\frac{1}{2} \left[\left( \frac{r_x-R_x}{l_x}\right)^2 + \left(\frac{r_y-R_y }{l_y}\right)^2 \right] }
\end{equation}
and that the momentum-space expansion coefficients of the Bloch states are  
\begin{equation}
 \Psi_\mathbf{Q}(\mathbf{G}) = 
    \frac{1}{\sqrt{N}} e^{-\frac{1}{2}\left[\mathbf{l}\cdot \mathbf{(\mathbf{Q}+\mathbf{G})}\right]^2}.
\end{equation}    
Here we defined the oscillator lengths $l_i \equiv [\hbar^2M/\kappa_i]^{1/4}$, $\mathbf{l} = (l_x, l_y)$. The normalization factors are given by $N_R = \pi l_xl_y$ and $N = \sum_{\mathbf{Q},\mathbf{G}} e^{-\left[\mathbf{l}\cdot \mathbf{(\mathbf{Q}+\mathbf{G})}\right]^2}$.  
Applying Eq. \ref{eq:matrixelement}, which assumes negligible mixing between bands, and projecting to
the lowest band yields the following diagonal (D) and off-diagonal (OD)
valley-space matrix elements for the exchange  Hamiltonian: 
\begin{eqnarray}
    \braket{J(\mathbf{Q})}_{D} &=& J\sum_\mathbf{G} |\Psi(\mathbf{Q}+\mathbf{G})|^2 |\mathbf{Q}+\mathbf{G}| \nonumber \\
    \braket{J(\mathbf{Q})}_{OD} &=& J\sum_\mathbf{G} |\Psi(\mathbf{Q}+\mathbf{G})|^2 |\mathbf{Q}+\mathbf{G}|e^{2i\theta_{\mathbf{Q}+\mathbf{G}}},
    \label{Eq:SHOJ}
\end{eqnarray}
The inter-valley coupling term is responsible for the gap and vanishes when $\alpha=1$ (no anisotropy.) 

We first focus on the spectrum at $\boldsymbol{\gamma}$ by evaluating these expressions at $\mathbf{Q}=0$.  To reveal the trends transparently, we assume that the reciprocal lattice sum can be truncated after the first shell, which is justified unless the oscillator lengths are short compared to the array lattice constant. In this case,
\begin{eqnarray}
    \braket{J(0)}_{D} &=& \frac{J}{N}\sum_{\mathbf{G}}|G|(e^{-(l_x G_{x})^2}+e^{-(l_yG_{y})^2}) \nonumber\\ &\approx& \frac{2JG}{N}(e^{-(l_x G)^2}+e^{-(l_yG)^2}) \nonumber \\
    \braket{J(0)}_{OD} &=& \frac{J}{N}\sum_{\mathbf{G}}|G|(e^{-(l_x G_{x})^2}-e^{-(l_yG_{y})^2}) \label{Eq:SHO} \\ &\approx& \frac{2JG}{N}(e^{-(l_x G)^2}-e^{-(l_yG)^2}) \nonumber
\end{eqnarray}
where $G\equiv|G_1|$. The inter-valley matrix element is non-zero when $l_x \ne l_y$, leading to a gap at $\gamma$. 

We now address the anisotropic band dispersion relevant to spreading of exciton clouds, which can be obtained by expanding Eq. \ref{Eq:SHOJ} to first order in $|\mathbf{Q}|$: 
\begin{eqnarray}   \braket{J(\mathbf{Q})}_{D} &\approx& \braket{J(0)}_{D} + 
|\Psi_{\mathbf{Q}}(\mathbf{G}=0)|^2 \, J|\mathbf{Q}| \,
\nonumber\\
\braket{J(\mathbf{Q})}_{OD} &\approx& \braket{J(0)}_{OD} +  |\Psi_{\mathbf{Q}}(\mathbf{G}=0)|^2\, J|\mathbf{Q}|e^{2i\theta_{\mathbf{Q}}}.
\label{Eq:SHOK}
\end{eqnarray}
Note that we have $\braket{J(0)}_{OD} > 0$ by convention, since we have assumed without loss of generality that $l_y = \alpha^{-1/4}l_x > l_x$.
The two lowest exciton bands are   
\begin{equation}
    E_\pm(\mathbf{Q}) \approx E^0+\braket{J(\mathbf{Q})}_{D} \pm |\braket{J(\mathbf{Q})}_{OD}|
    \end{equation}
where $E^0+E^1(\mathbf{Q})$ is the lowest band energy at $J=0$; since $E^1(\mathbf{Q})$ decreases exponentially with $a$, the dispersion is dominated by the exchange terms and we retain only the 
zero-point energy shift $E^0=\hbar(\omega_x+\omega_y)/2$. 
Then, the band dispersion to linear order in $|\mathbf{Q}|$ is given by
\begin{equation}
   E_\pm = E^0 + \frac{4JG}{N}e^{-(l_\pm G)^2}  + \frac{J}{N}|\mathbf{Q}| \left( 1 \pm  \cos(2 \theta_\mathbf{Q} )\right) + \mathcal{O}(Q^2)
    \label{Eq:SHOE}
\end{equation}
where $l_+,l_- \equiv l_x, l_y$.
It follows that the exciton dispersion in the lowest valley mode is linear except in the direction of strongest confinement - the $x$ direction in our case - and strongest in the direction of weakest confinement.  
For the higher valley mode the direction of strongest dispersion is the direction of strongest confinement. In addition, the dispersion of the two modes will match along the special directions $\theta_\mathbf{Q} = n\pi/2 $, $n\in \mathbb{Z}$. 

We emphasize that these expressions for dispersion of the lowest exciton bands neglect terms of order $|\mathbf{Q}|^2$ and higher.  When the bands are non-dispersive to linear order along some direction,
quadratic corrections are essential. In addition to quadratic exchange terms, mixing with higher SHO bands introduces level repulsion which makes a negative contribution to dispersion at order $|\mathbf{Q}|^2$ for the lowest energy valley-split bands, especially for the higher valley mode. 
An example of this trend can be discerned in the anisotropic square-lattice bands plotted in Fig. \ref{fig:K4Delta} (right).  A clearer example is provided by the quasi-1D striped geometry case 
discussed in the next section (see Fig. \ref{fig:StripeBands}, right). This downward dispersion 
has negative implications for the stability of the gapped valley modes as a two-level system addressable through optical excitation. In contrast, in the triangular-lattice case there is no direction
in which the upper valley-split mode has vanishing linear dispersion, and quadratic corrections
to not necessarily move the minimum of the upper band away from $\boldsymbol{\gamma}$.

\begin{figure}
    \centering \includegraphics[width=45mm]{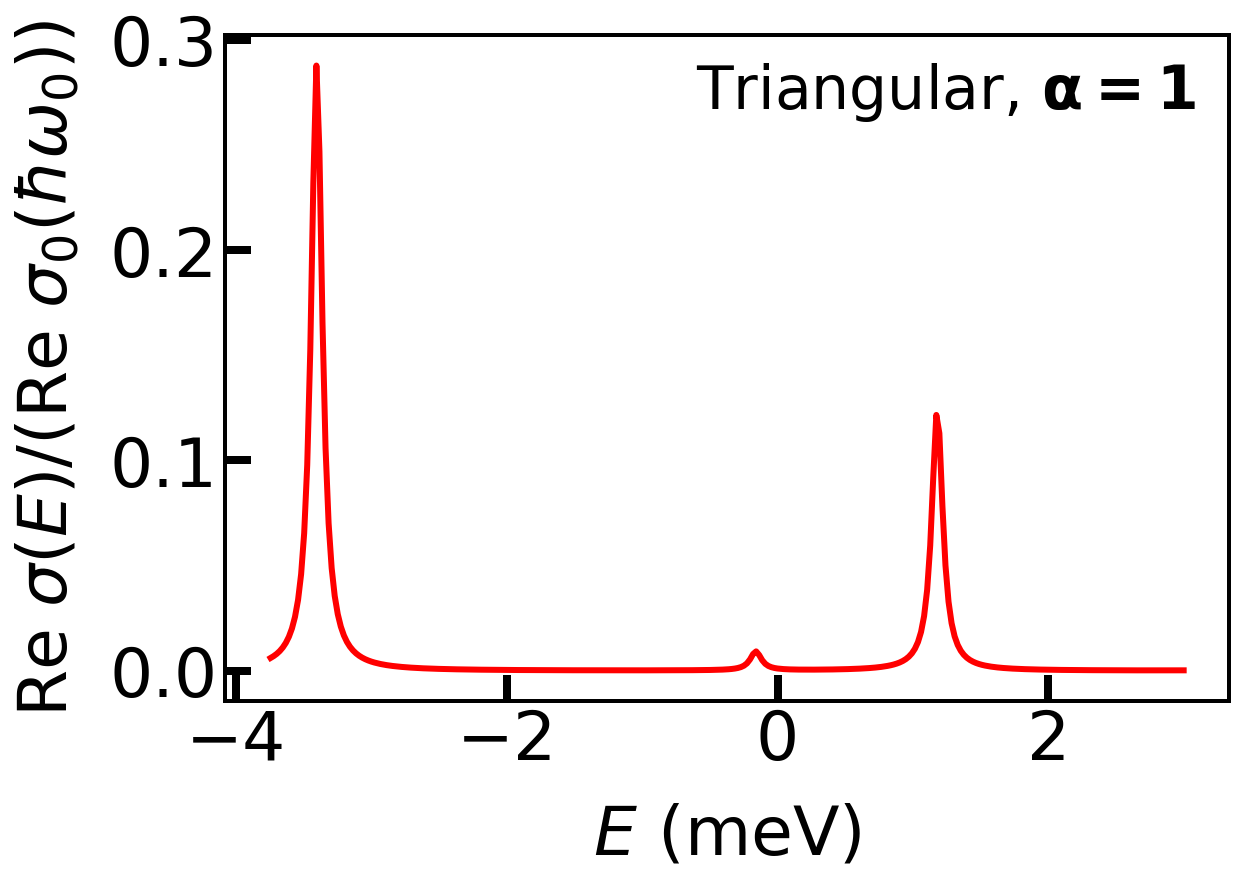} 
    \includegraphics[width=45mm]{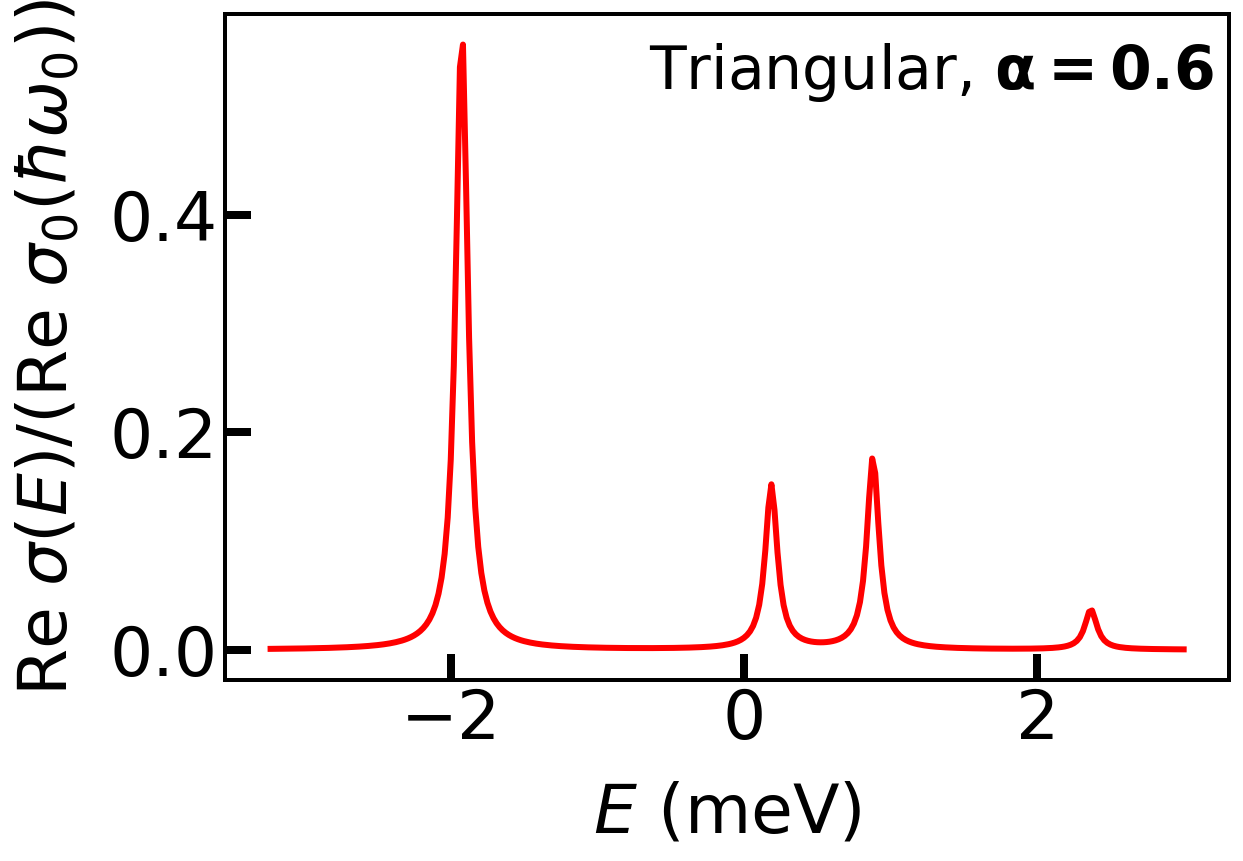}
    \includegraphics[width=45mm]{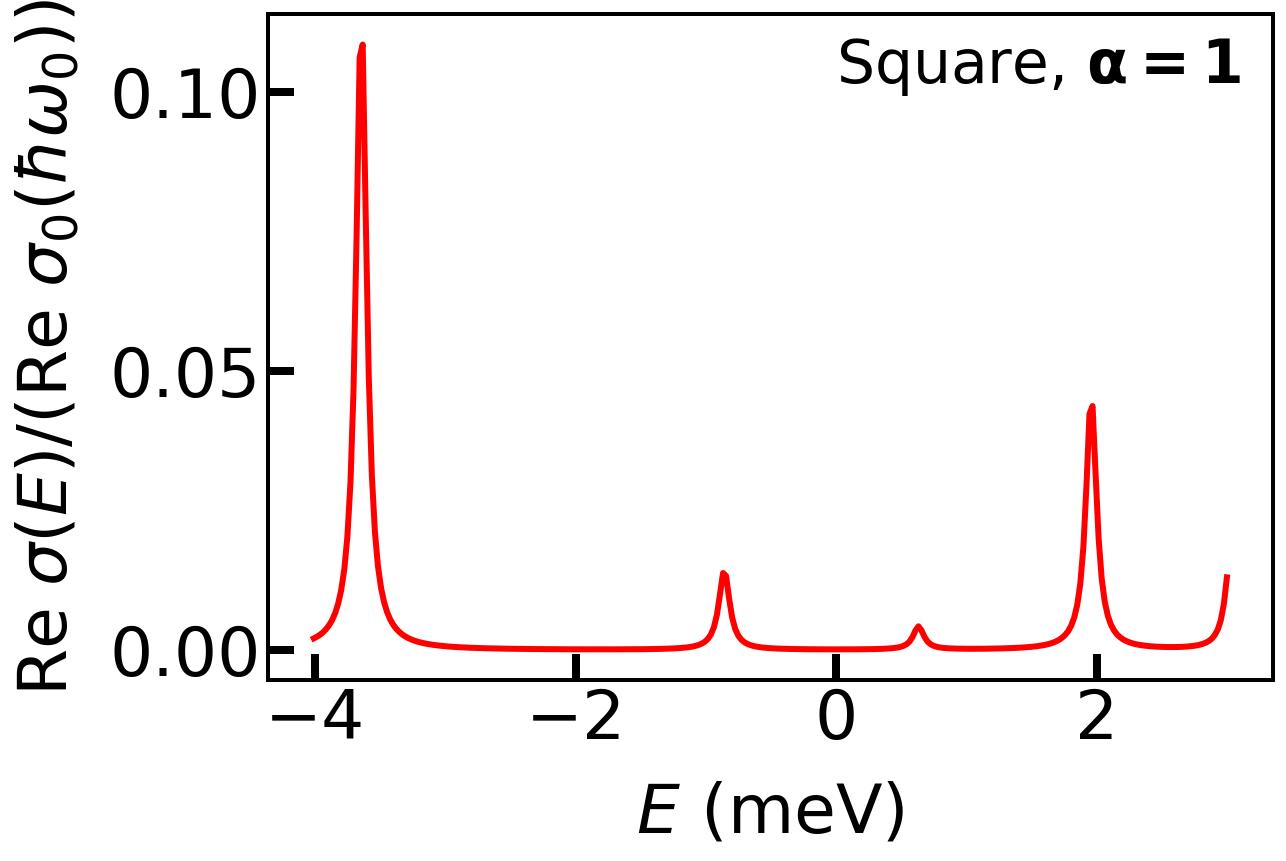} \includegraphics[width=44mm]{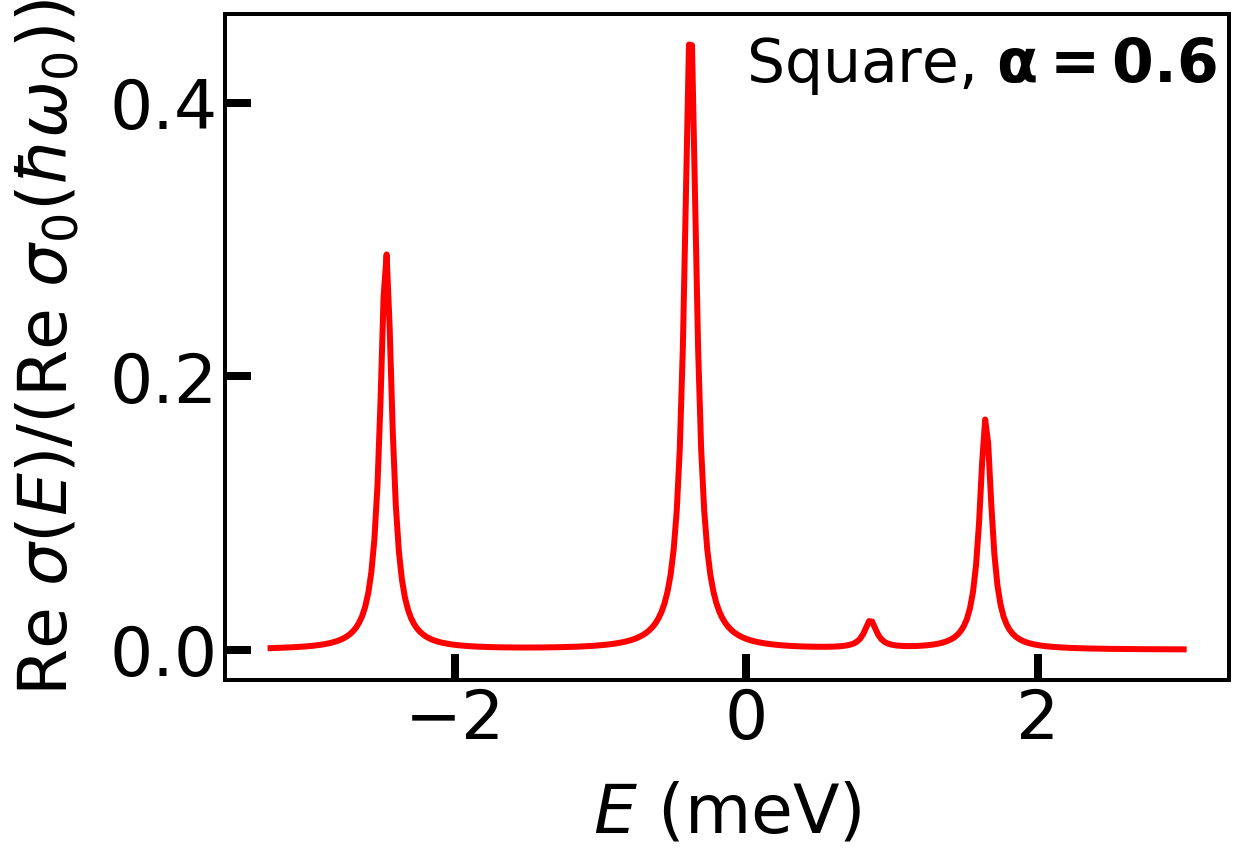}\hfill
    \caption{Optical responses in response to linearly polarized light, with broadening parameter $\eta = 0.05$ meV derived from the band structures in Fig. \ref{fig:K3Delta} (left) and Fig. \ref{fig:K4Delta} (right). Top: isotropy parameter $\alpha = 1$. Bottom: $\alpha = 0.6$.}
    \label{fig:OR}
\end{figure}

Finally, we can calculate the optical response of the exciton system from the exciton band states \cite{fogler2014high, animalu1970many}. The Kubo formula for the dissipative real part of the diagonal components of the conductivity tensor takes the following form when expressed in terms of exciton bands and COM wavefunctions:
\begin{equation}
    \begin{split}
    = \frac{\text{Re }\sigma_0(\omega_0) \eta}{2} \sum_{n}|\sum_{v} c^{n,v}_{\mathbf{G=0}}(\mathbf{Q=0})|^2 \Gamma(\omega-\omega_n) 
\end{split}
\end{equation}
with $\Gamma(\omega) = \eta/\left[\hbar^2\omega^2+\eta^2 \right]$ is a Lorentzian broadening 
function with broadening parameter $\eta$ and $\sigma_0(\omega_0)$ is the peak conductivity at the exciton resonance in the absence of electrostatic confinement.
Representative optical conductivities obtained from the band structures in this section are plotted in Fig. \ref{fig:OR}.  For generic light polarizations there is a peak in the optical absorption at all $\mathbf{Q}=0$ energies because all exciton Bloch states have $\mathbf{Q}=0$, $\mathbf{G}=0$ components; for the special case of  linearly polarized light, the weight in absorption of those exciton states that are antisymmetric in valley vanishes.

\section{Interdigitated gates}
\label{sec:three}
\begin{figure}
    \centering
    \includegraphics[width=45mm]{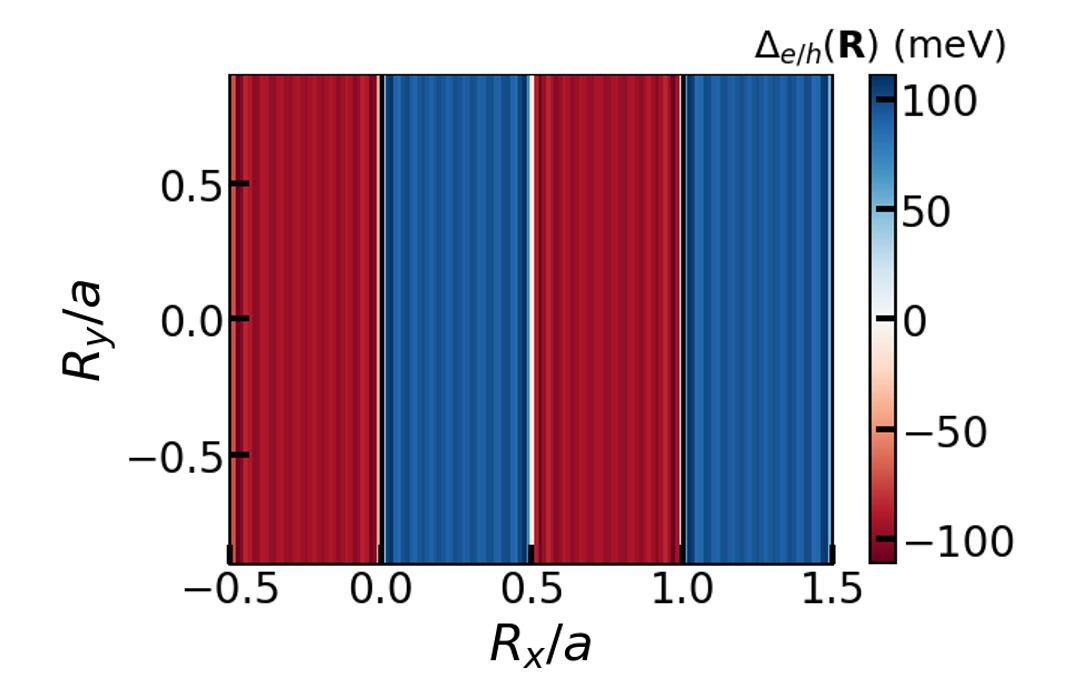}
    \hfill
    \includegraphics[width=44mm]{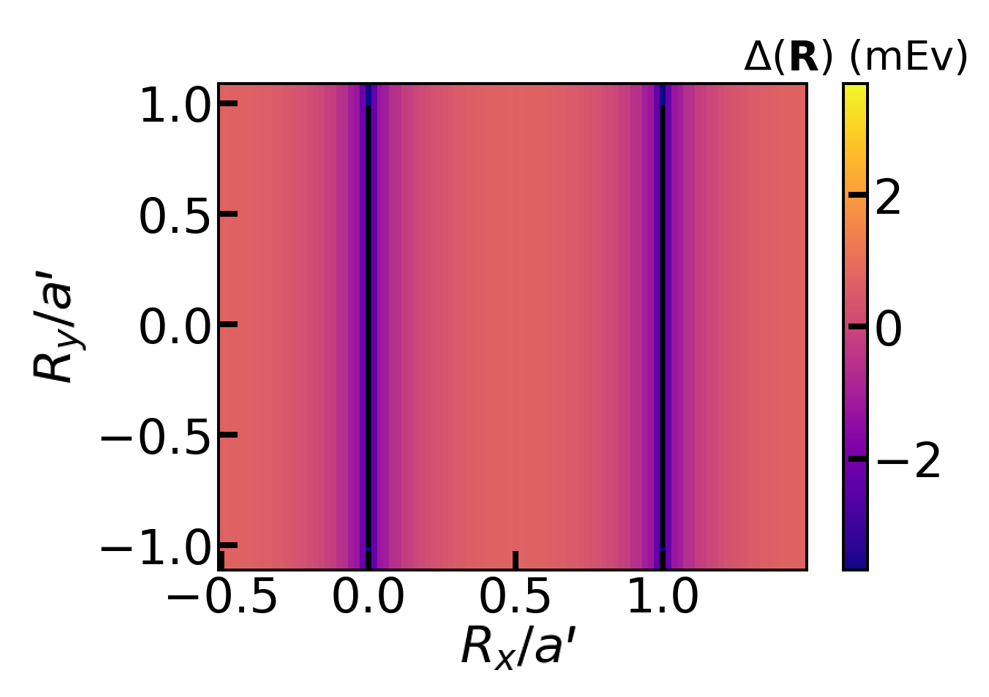}
    \caption{Gradients in the stripe-pattern electrostatic potential $\Delta_{e,h}$ (left) give rise to electric fields
    $\boldsymbol{E}$ which peak along the stripe boundaries and alternates directions. The resulting position-dependent Stark energy acts as an effective exciton potential $\Delta$ (right). The attractive domain walls have are separated by the domain width $a' \equiv a/2$.}
    \label{fig:StripeDelta}
\end{figure}

Motivated by the results in the previous section, we now consider the fully anisotropic limit
of a one-dimensional potential $\Delta_e$ formed by parallel stripe domains.  The $x$ and $y$ exciton
COM degrees-of-freedom then separate.  The electrostatic Hamiltonian is the sum of a one-dimensional
Kronig-Penney-like \cite{kronig1931quantum} one-dimensional periodic potential potential
with period $a'$ along the $x$ direction and a free particle Hamiltonian along $y$.  
This limit can be realized by fabricating interdigitated electrostatic gates \cite{hu2024quantum}. 
Sharp jumps in the  potential appear at the stripe edges which run along the $y$-direction at positions $x=0,a/2$ as illustrated shown in Fig.~\ref{fig:StripeDelta}. 
Because $\Delta_e$ is independent of the $y$ coordinate, the y-component of the
exciton momentum, $Q_y$, is a good quantum number. 
Note that the electrostatic potential has jumps at positions $x=0+na$ and 
jumps at positions $x=a/2+na$ which are inequivalent because they are associated 
with electric fields of opposite sign, but give rise to identical quadratic Stark effects.
The bosonic model therefore does not distinguish between the positions $x=0$ and $x=a/2$. 
For the bosonic model, we can set the lattice constant to $a'=a/2$ so that the DWs in the net exciton potential has period $a'$ as shown in Fig.~\ref{fig:StripeDelta}.

We study this limit numerically by Fourier expanding the exciton potential
\begin{equation}
   \Delta(x,y) = -2\sum^{2n_\text{max}}_{n} \Delta_{n} \cos{\left(\frac{2\pi nx}{a'}\right)},
\end{equation}
with $\Delta_n = V$ for all $n$ up to a cutoff.  
Consequently, the wavefunction can be separated into $x$ and $y$ components labeled by a crystal momentum $Q_x$ and a good vertical momentum $Q_y$:
\begin{align}
    \Psi_{Q_x;Q_y}(x,y) &= \psi_{Q_x}(x)\psi_{Q_y}(y) \\ &= \frac{1}{\sqrt{A}}e^{iQ_xx}\sum_{j\in \mathbb{Z}}\left( c_{Q_x}(G_j)e^{iG_jx}\right)e^{iQ_yy}
\end{align}
where $G_j = 2j\pi/a$ are the RL vectors for the periodic $x-$direction, $c_{Q_x}(G_j)$ are the $\psi_{Q_x}$ normalized coefficients in this RL plane-wave basis, and $A$ is the system area. In this section, we use the parameters $V = 0.5$ meV, $a' = 50$ nm, $J = 0.4$ eV$\cdot$nm , and truncate the Fourier sum at $n_\text{max}=22$ to obtain convergence of the band energies with a precision of $\lesssim 10^{-3}$ meV.

\begin{figure}
    \centering \includegraphics[width=45mm]{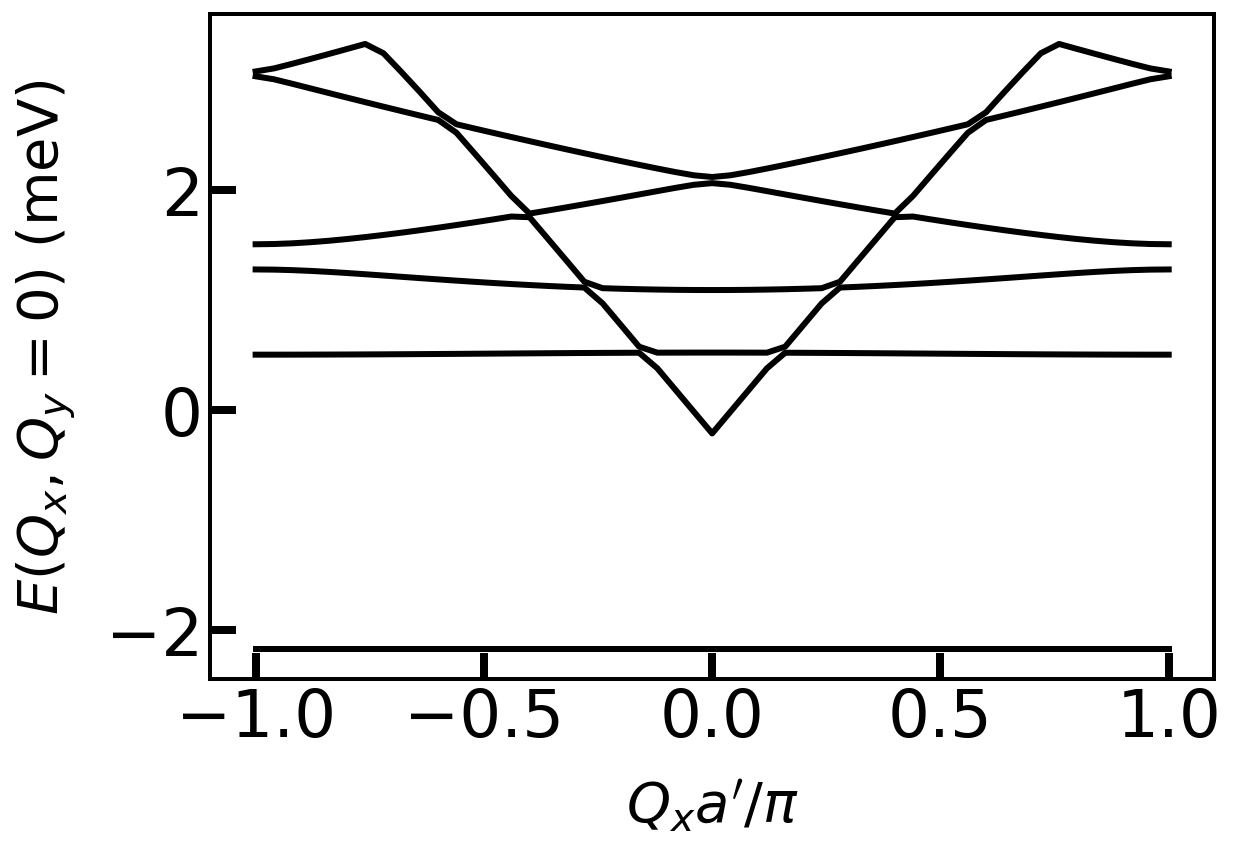} \includegraphics[width=45mm]{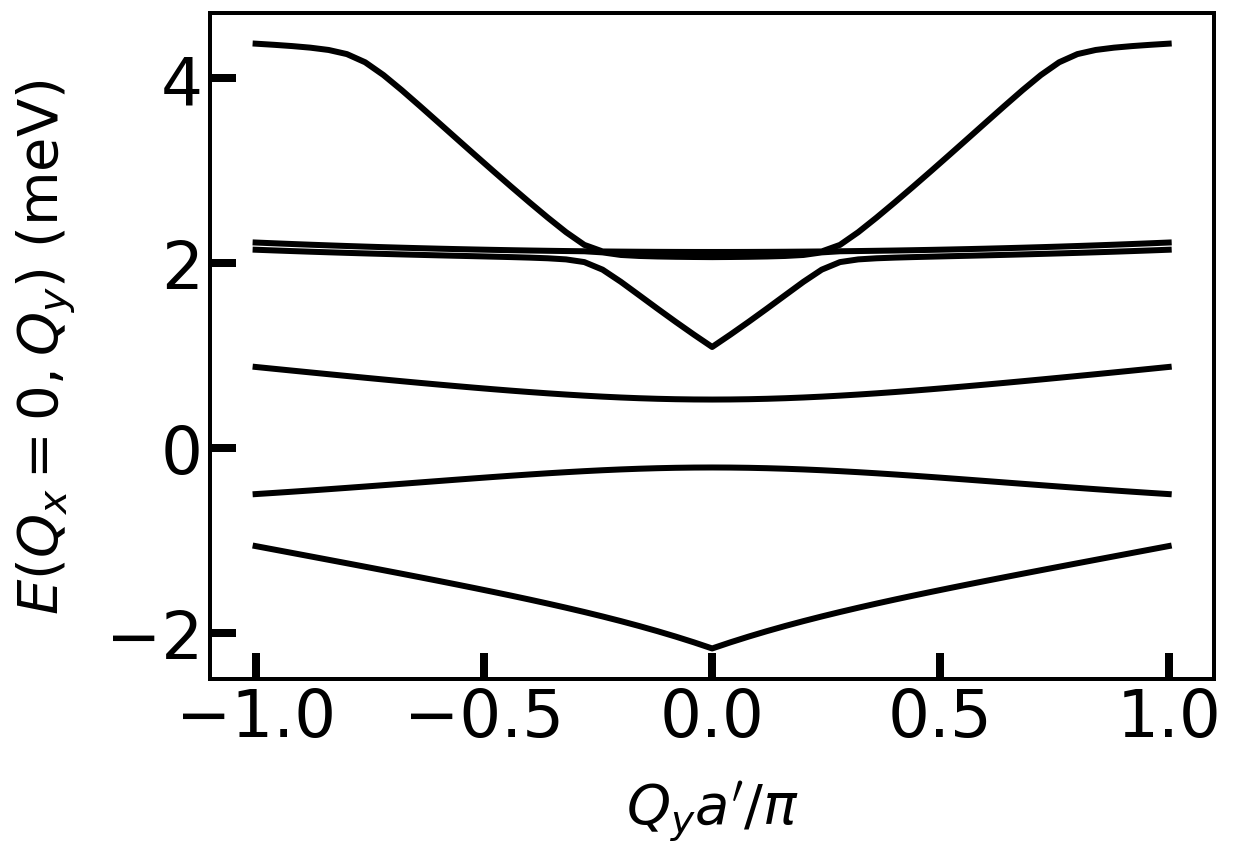} 
    \caption{
    Exciton bands for a quasi-1D harmonic potential, plotted along the directions $Q_x = 0$ (left) and 
    $Q_y = 0$ (right). The potential gradients are maximized along 
    an array of parallel lines with period $a' = 50$ nm.} 
    \label{fig:StripeBands}
\end{figure}
The resulting exciton bands are illustrated in Fig.~\ref{fig:StripeBands} as a function of crystal wavevector $Q_x \in (-\pi/a',\pi/a') \equiv (-m,m)$ and unfolded momentum $Q_y$. 
The plot range for $Q_y$ is truncated to $\pm \pi/a'$ to simplify comparison of $Q_x$ and $Q_y$ dependences.  As in the more modestly asymmetric cases discussed previously, 
exchange induces a splitting of the $\mathbf{Q}=0$ exciton bands and 
linear exciton dispersion except in special directions.
We dub the two modes ``exchange-parallel" and ``exchange-perpendicular" 
to emphasize their rapid-dispersion directions relative to the weak confinement direction, which in our case is the $y-$direction. These bands can also be plotted using the microscopic BZ $[-\pi/a,\pi/a] = [-\pi/(2a'),\pi/(2a')]$, which amounts to folding the bands in half. In that case, the bands come in pairwise degenerate at the BZ boundaries; this can be understood by analogy to a two-sublattice tight-binding model, chiefly the equal-hopping limit of the Su-Schrieffer-Heger model of polyacetylene \cite{heeger1988solitons}. 
However, it is simpler for us to analyze the striped-domain case using the machinery developed in the previous sections for anisotropic potentials. In fact, the fully-anisotropic $\alpha = 0$ limit of the SHO toy model produces a similar exciton potential with minima along 1D parallel lines; the only difference between the two cases then becomes the sharpness of the attractive DWs, a quantitative property determined by the number of non-zero harmonics in the expansion of the potential. That is, this comparison becomes exact within the first-shell harmonic approximation. Therefore, we consider the limit $\alpha = 0$ of Eq. \ref{Eq:SHOJ}, which implies $l_y \longrightarrow{} +\infty$ and hence
\begin{eqnarray}
\braket{J(0)}^{\tau, \tau'} &=& \frac{2J}{N}\sum_\mathbf{G}|\mathbf{G}|(e^{-(l_xG_x)^2}+(-1)^{\delta_{\tau,\tau'}} e^{-(l_yG_y)^2}) \nonumber \\
 &\longrightarrow& \frac{2J}{N}\sum_\mathbf{G}|\mathbf{G}|e^{-(l_xG_x)^2}\approx \frac{2J}{N}G e^{-(l_xG)^2} 
\end{eqnarray}
where $G=2\pi/a'$ now and we applied the first-shell truncation to approximate the sum in the last inequality.
Therefore, we find that the matrix elements $\braket{J(0)}$ become valley-independent in this limit, with the intravalley component being minimized and the intervalley one (as well as the exchange gap) maximized. 
Then, the mode dispersions are given in analogy to Eq. \ref{Eq:SHOE} by
\begin{equation}
   E_+ - E_- =  \frac{4JG}{N} e^{-(l_x G)^2} + \frac{2J}{N}|\mathbf{Q}| \  \cos(2 \theta_\mathbf{Q} )e^{-(l_x Q_x)^2}.
\end{equation}
These expressions are consistent with the observation that confinement anisotropy correlates positively and strongly with the size of the valley gap and with the linear-quadratic dispersion anisotropy as shown by our numerical results. \\

\section{Conclusions}
\label{sec:five}
The excitons in TMDs and other van-del-Waals systems are often characterized by a valley label taking values at the $K,K'$ corners of the Brillouin zone. Despite time-reversal and $C_n$ symmetry, the valley degeneracy is lifted everywhere in the BZ except at the $\boldsymbol{\gamma}$ point by the e-h exchange interaction, which includes a valley-mixing term and grows linearly with the magnitude of the exciton CM momentum $\mathbf{Q}$. Previous theoretical work \cite{wu2017topological, glazov2025long-range} focused on time-reversal as the symmetry protecting the band touching at $\boldsymbol{\gamma}$, and then proposed the application of a Zeeman field to engineer isolated bands compatible with non-trivial topology or exciton superfluidity. In contrast, we have shown here that the key ingredient for the exchange-induced gap is breaking of in-plane rotational symmetry, so that anisotropic strain or gating patterns can be leveraged to open an exchange gap and modulate the exciton valley state. In this case, the confinement geometry dictates a particular direction of the exciton valley pseudospin at $\boldsymbol{\gamma}$. Moreover, the resulting exciton bands present anisotropic dispersions with mixed linear and quadratic character. For the square-lattice and striped-domain geometries, we showed that the higher (or 'exchange-perpendicular) valley mode acquires downwards quadratic dispersion along the weak confinement direction, which makes this valley-excited exciton unstable. In contrast, the dispersion of this mode in the triangular-lattice case is predominantly linear with a stable minimum at $\boldsymbol{\gamma}$.
Moreover, because the lowest-energy exciton band is linearly-dispersive (expect for a set of angles with zero dimension), Bose-Einstein condensation of excitons into a true superfluid state with long-range-off-diagonal-order is allowed \cite{kapitza1938viscosity, allen1938flow, blatt1962bose}. The linear dispersion suppresses thermal fluctuations, permitting
macroscopic occupation of the lowest-energy $\mathbf{Q}=0$ exciton.
The role of exchange interactions in favoring long-range-order is similar to its role in the case of interlayer excitons in a magnetic field analyzed in Refs. \cite{li2017excitonic, glazov2025long-range}. We leave the detailed study of anisotropy-induced off-diagonal-long-range-order for future work. Our results offer insight into the properties of trapped excitons in gated and stacked TMD systems, as well as potential applications for valleytronics and coherent light emission.
\section{Acknowledgements}
Work supported by U.S. Department of Energy, Office of Basic Energy Sciences (DE-SC0021984). 
J.M.T and A.H.M. are thankful for discussions with D. Kim, B. Zou, N. Wei, Y. Zeng and X. Chen.
\bibliography{excitonbands}

\end{document}